\documentclass[aps,twocolumn,showpacs,superscriptaddress,eqsecnum]{revtex4}

\usepackage{epsfig}
\usepackage{graphicx}
\usepackage{amsfonts}
\usepackage[figuresright]{rotating}  
\usepackage{amssymb}
\usepackage{amsmath}
\usepackage{psfrag}
\usepackage{subfigure}

\def\avg#1{\langle#1\rangle}

\def\tr{\mbox{tr}}
\def\be{\begin{equation}}       
\def\ee{\end{equation}}
\def\bea{\begin{eqnarray}}      
\def\eea{\end{eqnarray}}

\def\nn{\nonumber}
\def\pp{\parallel}

\begin{document}
\title{Fermi liquid instabilities in the spin channel}
\author{Congjun Wu}
\affiliation{Kavli Institute for Theoretical Physics, University
of California, Santa  Barbara, California 93106-4030}
\author{Kai Sun}
\affiliation{Department of Physics, University of Illinois at Urbana-Champaign,
Urbana, Illinois 61801-3080}
\author{Eduardo Fradkin}
\affiliation{Department of Physics, University of Illinois at Urbana-Champaign,
Urbana, Illinois 61801-3080}
\author{Shou-Cheng Zhang}
\affiliation{Department of Physics, McCullough Building, Stanford
University, Stanford, California 94305-4045}

\begin{abstract}
We study the Fermi surface instabilities of the Pomeranchuk
type in the spin triplet channel with high orbital partial waves  
($F_{l}^a ~(l>0)$).
The ordered phases are classified into two classes, dubbed the $\alpha$ 
and $\beta$-phases by analogy to the superfluid  $^3$He-A and B-phases.
The Fermi surfaces in the $\alpha$-phases exhibit spontaneous anisotropic distortions,
while those in the $\beta$-phases remain circular or spherical
with topologically non-trivial spin configurations in momentum space.
In the $\alpha$-phase, the Goldstone modes in the density channel
exhibit anisotropic overdamping.
The Goldstone modes in the spin channel have nearly isotropic underdamped 
dispersion relation at small propagating wavevectors.
Due to the coupling to the Goldstone modes, the spin wave
spectrum develops resonance peaks in both the $\alpha$ and $\beta$-phases,
which can be detected in inelastic neutron scattering experiments.
In the $p$-wave channel $\beta$-phase, a chiral ground state inhomogeneity   
is spontaneously generated due to a Lifshitz-like instability
in the originally nonchiral systems.
Possible experiments to detect these phases are discussed.
\end{abstract}
\date{\today}
\maketitle 

\section{Introduction}
\label{sec:intro}

The Landau theory of the Fermi liquid is one of the most successful theories
of condensed matter physics~\cite{leggett1975,baym1991}.  
It describes a stable phase of dense interacting fermionic systems, 
a {\em Fermi liquid} (FL).
Fermi liquid theory is the foundation of our understanding of 
conventional, weakly correlated, metallic systems.
Its central assumption is the existence of well-defined
fermionic quasi-particles, single particle fermionic excitations 
which exist as long-lived states at very low energies, close enough
to the Fermi surface. 
In the Landau theory, the interactions among quasi-particles are captured
by a few Landau parameters $F^{s,a}_{l}$,  where $l$ 
denotes the orbital angular momentum partial wave channel, and $s,a$ denote 
spin singlet and triplet channels, respectively.
Physical quantities,
such as the spin susceptibility, and properties of collective excitations,
such as the dispersion relation of zero sound collective modes,
acquire significant but finite renormalizations due to the Landau interactions.
In the FL phase, except for these finite renormalizations, the effects of
the interactions become negligible at asymptotically low energies. 
It has, however, long been known that the stability of the FL requires 
that the Landau parameters cannot be too negative, $F_l^{s,a}> -(2l+1)$, 
a result first derived by Pomeranchuk~\cite{pomeranchuk1959}.
The most familiar of these Pomeranchuk instabilities are found
in the $s$-wave channel:
the Stoner ferromagnetism at $F^a_0<-1$ and phase
separation at $F^s_0<-1$.

It has been realized quite recently that when these bounds are violated in
a channel with a non-vanishing angular momentum, there is a ground state 
instability in the particle-hole, spin singlet, channel leading to a
spontaneous distortion of the Fermi surface. 
This is a quantum phase transition to a uniform but anisotropic liquid
phase of the fermionic system~\cite{oganesyan2001}. 
In such a phase the electron fluid behaves, from the point of view of its
symmetries and of their breaking, very much like an electronic analog of 
liquid crystal phases~\cite{degennes1993,chaikin1995}. 
The charge nematic is the simplest example of such electronic liquid 
crystal phases, a concept introduced in Ref.~\cite{Kivelson98} to 
describe the complex phases of strongly correlated systems such as
doped Mott insulators. The charge nematic phase has also been 
suggested to exist in the high $T_c$ materials 
near the melting of the (smectic) stripe 
phases~\cite{Kivelson98,kivelson2003}, and in quantum Hall systems 
in nearly half-filled Landau levels~\cite{fradkin99,fradkin2000}.  
Experimentally, the charge nematic phase 
has been found in ultra-high mobility two-dimensional electron gases 
(2DEG) in  AlAs-GaAs heterostructures and quantum wells in large 
magnetic fields, in nearly half filled Landau levels for $N\geq 2$ at 
very low temperatures~\cite{lilly1999,cooper02,du1999,fradkin2000}. 
Strong evidence for a charge nematic phase has been found quite recently
near the metamagnetic transition of the ultra-clean samples of the 
bilayer ruthenate $Sr_3Ru_2O_7$~\cite{grigera2004,mackenzie-comm}.

Electronic liquid crystal phases can also be realized as 
Pomeranchuk instabilities in the particle-hole channel with non-zero
angular momentum. 
This point of view has been the focus of much recent work, both 
in continuum system~\cite{oganesyan2001,barci2003,lawler2006,Nilsson2006,quintanilla2006} as well as in lattice systems~\cite{halboth2000,dellanna2006,kee2003,khakvine04,yamase05}, in which case these quantum phase transition involve the spontaneous breaking of the point symmetry group of the underlying lattice. The 2D quantum nematic Fermi fluid phase is an instability in the $d$-wave ($l=2$) channel exhibiting
a spontaneous elliptical distortion of the Fermi surface~\cite{oganesyan2001}.
In all cases, these instabilities typically result in anisotropic Fermi 
surface distortions, sometimes with a change of the topology of the Fermi surface. Nematic phases also occur in strong coupling regimes of strongly correlated systems such as the Emery model of the high temperature superconductors~\cite{geballe04}. 
If lattice effects are ignored, the nematic state has Goldstone modes can be viewed as the rotation 
of the distorted Fermi surface, {\it i.e.\/}, the soft Fermi surface fluctuations.
Within a random phase approximation (RPA) approach~\cite{oganesyan2001}, confirmed by a non-perturbative high dimensional bosonization treatment
\cite{lawler2006},
the Goldstone modes were shown to be overdamped in almost all the propagating directions, 
except along the high symmetry axes of the distorted Fermi surface.
The Goldstone mode couples strongly to electrons, giving rise to a non-Fermi liquid
behavior throughout the nematic Fermi fluid phase~\cite{oganesyan2001,lawler06b}: in perturbation theory, the imaginary part of the electron self-energy is found to be proportional
to $\omega^{\frac{2}{3}}$ on most of the Fermi surface, except along four ``nodal'' directions, leading to the breakdown of the
quasi-particle picture. Away from the quantum critical point, this effect is suppressed by lattice effects but it is recovered at quantum criticality~\cite{dellanna2006} and at high temperatures (if the lattice pinning effects are weak). Beyond perturbation theory~\cite{lawler06b} this effect leads to a form of `local quantum criticality'.      

Richer behaviors still can be found in Pomeranchuk
instabilities in  the spin triplet and  $l > 0$ angular momentum channels
\cite{hirsch1990,hirsch1990a,gorkov1992,oganesyan2001,
kivelson2003,wu2004,varma2005,varma2006,Kee2005,Honerkamp2005}.
Typically, these class of instabilities break both the spacial (orbital)
and spin $SU(2)$ rotation symmetries.
A $p$-wave channel instability was studied by Hirsch~\cite{hirsch1990,hirsch1990a}.
The Fermi surfaces of spin up and down components
in such a state shift along opposite directions.
The $p$-wave channel instability was also proposed by Varma
as a candidate for the hidden order appearing in the heavy fermion compound
URu$_2$Si$_2$~\cite{varma2005,varma2006} below 17 $K$.
Gor'kov and Sokol studied the non-N\'eel spin orderings in itinerant
systems, and showed their relation to the Pomeranchuk 
instability~\cite{gorkov1992}.
Oganesyan 
and coworkers \cite{oganesyan2001,kivelson2003} proposed the existence of a  nematic-spin-nematic phase, a nematic state in both real space and in the internal spin space, where
the two Fermi surfaces of up and down spins are spontaneously distorted into two orthogonal ellipses. Podolsky and Demler\cite{Podolsky2005} considered a spin-nematic phase as arising from the melting of a stripe phase; this phase can also be a nematic-spin-nematic if it retained a broken rotational symmetry, as expected from the melting of a stripe (or smectic) phase\cite{Kivelson98,Zaanen2003}. Kee and Kim~\cite{Kee2005} have suggested a state to explain the behavior of $Sr_3Ru_2O_7$ which can be shown to be a lattice version of a partially polarized nematic-spin-nematic state.

Interestingly, the spin triplet Pomeranchuk instabilities can also occur without
breaking rotational invariance in real space, and keep the symmetries of the 
undistorted Fermi surface.
Wu {\it et al.} \cite{wu2004} showed that a state of this sort can exist in the $p$-wave channel,
with the distortions affecting only the spin channel.
In this state, there are two Fermi surfaces with different volumes,
as in ferromagnetic systems.
Although both the spatial and spin rotation symmetries are broken,
a combined spin-orbit rotation keeps the system invariant, {\it i.e.\/},
the total angular momentum is conserved.
The broken symmetry in such a state is the relative spin-orbit
symmetry which was first proposed by Leggett in the context of
superfluid $^3$He-B phase \cite{leggett1975}.
In fact, it is a particle-hole channel analog of the superfluid $^3$He-B
phase where the pairing gap function is isotropic over the Fermi surface.
The phases with anisotropic Fermi surfaces distortions
are the analog of the $^3$He-A phase where the gap function
is anisotropic.
The two possibilities of keeping or breaking the shape of Fermi surfaces
are dubbed $\beta$ and $\alpha$-phases respectively,
by analogy with  B and A superfluid phases in $^3$He systems. An important difference is, however, that while in the A and B phases of superfluid $^3$He all the Fermi surface is gapped, up to a set of measure zero of nodal points in the A phase, in the $\beta$ and $\alpha$-phases of spin triplet Pomeranchuk systems no gap in the fermionic spectrum ever develops.
 
An important common feature of the $\alpha$ and $\beta$-phases is 
the dynamical appearance of effective spin-orbit (SO) couplings, reflecting
the fact that in these phases spin and orbit degrees of freedom become 
entangled \cite{warning}.
Conventionally,  in atomic physics, the SO coupling originates from leading
order relativistic corrections to the Schr\"odinger-Pauli equation. 
As such, the standard SO effects in many-body systems have an inherently 
single-particle origin, and are unrelated to many-body correlation effects.
The Pomeranchuk instabilities involving spin we are discussing here thus 
provide a new mechanism to generate effective SO couplings through
phase transitions in a many-body non-relativistic systems.
In the 2D $\beta$-phase, both Rashba and Dresselhaus-like SO couplings
can be generated.
In the $\alpha$-phase, the resulting SO coupling  can be
considered as a mixture of Rashba and Dresselhaus with equal
coupling strength.
Such SO coupling systems could in principle be realized in 2D semiconductor
materials leading to interesting new effects.
For instance, a hidden $SU(2)$ symmetry in such systems was found to give rise to
a long lived spin spiral excitation with
the characteristic wavevector relating the two Fermi surfaces
together \cite{bernevig2006}.
Recently, many proposals have been suggested to employ SO coupling
in semiconductor materials to generate spin current through
electric fields.
The theoretical prediction of this ``intrinsic spin Hall effect"
\cite{murakami2003,sinova2004} has stimulated tremendous
research activity both theoretical and experimental
\cite{kato2004,wunderlich2005}.
Thus, Pomeranchuk instabilities in the spin channel may have a potential
application to the field of spintronics.

A systematic description of the high partial-wave channel 
Pomeranchuk instabilities involving spin is still lacking in the
literature.
In this paper, we investigate this problem for arbitrary
orbital partial wave channels in two dimensions (2D),  and for simplicity only in the $p$-wave
channel in three dimensions (3D).
We use a microscopic model to construct a general Ginzburg-Landau (GL) free energy
to describe these instabilities showing that 
the structure of the $\alpha$ and $\beta$-phases 
are general for arbitrary values of $l$.
The $\alpha$-phases exhibit anisotropic relative distortions 
for two Fermi surfaces as presented in previous publications.
We also investigate the allowed topological excitations (textures) of these phases and find that 
a half-quantum vortex-like defect in real space, combined with spin-orbit distortions.
The $\beta$-phases at $l\ge 2$ also have a vortex
configuration in momentum space with winding numbers $\pm l$
which are equivalent to each other by a symmetry transformation. 

We  study the collective modes in critical regime
and in the ordered phases at zero temperature.
At the quantum critical point, as in the cases previously studied, the theory has dynamic critical
exponent $z=3$ for all values of the orbital angular momentum $l$.
In the anisotropic $\alpha$-phase, the Goldstone modes  can be classified into
density and spin channel modes, respectively.
The density channel Goldstone mode exhibits anisotropic overdamping
in almost all the propagating directions.
In contrast, the spin channel Goldstone modes show nearly isotropic
underdamped dispersion relation at small propagating wavevectors.
In the $\beta$-phase, the Goldstone modes are relative spin-orbit
rotations which have linear dispersion relations at $l\ge 2$,
in contrast to the quadratic spin-wave dispersion in the ferromagnet.
Both the Goldstone modes (spin channel) in the $\alpha$-phases and
the relative spin-orbit Goldstone modes in the $\beta$-phases 
couple to spin excitations in the ordered phases.
Thus the spin wave spectra develop characteristic resonance peaks 
observable in neutron scattering experiments,
which are absent in the normal phase.

The $p$-wave channel Pomeranchuk instability involving spin
is special because the Ginzburg-Landau (GL) free energy 
contains a cubic term of order parameters with a linear
spatial derivative satisfying all the symmetry requirement.
Such a term is not allowed in other channels with $l\ne 1$,
including the ferromagnetic instability and the 
Pomeranchuk instabilities in the density channel.
This term does not play an important role in the $\alpha$-phase.
But in the $\beta$-phase, it induces a chiral inhomogeneous 
ground state configuration leading to a Lifshitz-like instability 
in this originally nonchiral system.
In other words, a Dzyaloshinskii-Moriya type interaction for the
Goldstone modes is generated in the $\beta$-phase.
The effect bears some similarity to the helimagnet \cite{binz2006}
and  chiral liquid crystal \cite{chaikin1995} except that the parity 
is explicitly broken there but not here.
The spiral pattern of the ground state order parameter is
determined.

The paper is organized as follows:
In Section \ref{sec:model}, we construct the model Hamiltonian
for the Pomeranchuk instabilities with spin.
In Section \ref{sec:GLfree}, we present the Ginzburg-Landau
free energy analysis to determine the allowed ground states.
In Section \ref{sec:meanfield}, we discuss the results of the mean field theory. In Section \ref{sec:topology} we discuss the topology of the broken symmetry states and classify the topological defects.
In Section \ref{sec:critical}, we calculate the collective modes
at the critical point at zero temperature.
In Section \ref{sec:alpha} and Section \ref{sec:beta}, 
we investigate the Goldstone modes in the $\alpha$ and $\beta$-phases
respectively, and study the spontaneous Lifshitz instability
in the $\beta$-phase at $l=1$.
In Section \ref{sec:magnetic}, we present the magnetic field
effects.
In Section \ref{sec:spincurrent}, we show that in the
$\alpha$ and $\beta$-phases in the quadrupolar  channel
($l=2$), a spin current can be induced
when a charge current flows through the system.
In Section \ref{sec:experiment}, we discuss the possible
experimental evidence for Pomeranchuk instabilities involving spins.
We summarize the results of this paper at Section \ref{sec:conclusion}.
Details of our calculations are presented in two appendices. 
In Appendix \ref{app:landau-parameters} we specify our conventions 
for Landau parameters and in Appendix \ref{app:gsspinealpha}
we give details of the Goldstone modes for spin oscillations in 
the $\alpha$-phase.

\section{Model Landau Hamiltonian}
\label{sec:model}

We begin with the model Hamiltonian describing the Pomeranchuk
instability in the $F^a_{l} ~(l\ge 1)$ channel in 2D. 
Later on in the paper we will adapt this scheme to discuss the 3D case 
which is more complex. This model, and the related model for the spin-singlet sector of Ref.~\cite{oganesyan2001} on which it is inspired, has the same structure as the effective Hamiltonian for the Landau theory of a FL.
The corresponding order parameters can be defined through the matrix form as
\bea
Q^{\mu b}(r)&=&\psi^\dagger_\alpha(r) \sigma^\mu_{\alpha\beta} 
g_{l,b} (-i\hat \nabla) \psi_\beta(r)
\label{eq:order-parameters}
\eea
where the Greek indices $\mu$ denote the $x,y,z$ directions in the spin space, and in 2D 
the latin indices $b=1,2$  denote the two orbital components, and $\alpha,\beta=\uparrow,\downarrow$ label the two spin projections. (Hereafter, repeated indices are summed over.)
In 2D, the operators $g_{l,1}\pm ig_{l,2}$, which carry the azimuthal angular momentum quantum number $L_z=\pm l$, are given by
\bea
g_{l,1}(-i \hat \nabla) \pm i g_{l,2}(-i \hat \nabla)
= (-i)^l (\hat \nabla_x \pm i \hat \nabla_y)^l,
\label{eq:g12}
\eea
where the operator $\hat \nabla^a$ is defined as
$\vec \nabla^a/|\nabla|$. The 3D counterpart of these expressions can be written in  terms of spherical harmonic functions. Thus, in 3D the latin labels take $2l+1$ values. For the moment, and for simplicity, we will discuss first the 2D case.

In momentum space ({\it i.e.\/} a Fourier transform) we can write the operators of Eq. \eqref{eq:g12} in the form
$g_{l,1}(\vec k)=\cos l\theta_k$ and $g_{l,2}(\vec k)=\sin l\theta_k$,
where $\theta_k$ is the azimuthal angle of $\vec k$ in the 2D plane.
In momentum space, $Q^{\mu b}(\vec q)$ is defined as
\bea
Q^{\mu b}(\vec q)= \sum_{\vec k}\psi^\dagger_\alpha(\vec k+ \frac{\vec q}{2})
~\sigma^\mu_{\alpha\beta}~ g_{l,b} (\vec k) ~
\psi_\beta(\vec k- \frac{\vec q}{2}).
\label{eq:Q-momentum}
\eea
It satisfies $Q^{\mu b}(-\vec q)=Q^{\mu, b,*} (\vec q)$,
thus $Q^{\mu b}(\vec r)$ is real.

We generalize the Hamiltonian studied in Ref.
\cite{oganesyan2001,wu2004} to the $F^a_l$
channel with arbitrary values of $l$ as
\bea
H&=&\int d^2{\vec r} ~\psi^\dagger_\alpha({\vec r})
(\epsilon(\vec{\nabla})
-\mu) \psi_\alpha({\vec r}) \nonumber \\
&+& \frac{1}{2}
\int d^d{\vec r} d^d{\vec  r}^\prime ~f^a_l({\vec  r- \vec r^\prime)}
\sum_{\mu b} 
{\hat Q}^{\mu b} ({\vec r}){\hat Q}^{\mu b} ({\vec r^\prime}),
\nn \\
\label{eq:Ham}
\eea
where $\mu$ is the chemical potential.
For later convenience, we include the non-linear momentum dependence 
in the single particle spectrum up to the cubic level as
\bea
\epsilon(\vec{k})=v_F \Delta k (1+ a (\Delta k/k_F)
+b(\Delta k/k_F)^2
+\ldots)
\label{eq:dispersion}
\eea
with $\Delta k= k-k_F$. Here $v_F$ and $k_F$ are the Fermi velocity 
and the magnitude of the Fermi wave vector in the FL.

The Fourier transform of the Landau interaction function $f(r)$ is 
\bea
f(\vec q) =\int d \vec r e^{i\vec q \vec r}
f^a_l(r)= \frac{f^a_l}{ 1+\kappa |f^a_l| q^2},
\eea
and the dimensionless Landau parameters are defined as
\bea
F_{l}^a=N(0) f_l^a (q=0)
\eea
with $N(0)$ the density of states at the Fermi energy.
This Hamiltonian possesses the symmetry of the direct product of
$SO_L(2) \otimes SO_S(3)$ in the orbit and spin channels.

The LP instability occurs at $ F_l^a<-2$ at $l\ge 1$
in two dimensions.
For the general values of $l$, we represent the order parameter
by a $3\times 2$ matrix 
\bea
n^{\mu,b}&=&|f^a_1|\int \frac{d^2 \vec k}{(2\pi)^2}
\avg{\psi^\dagger_\alpha (k) \sigma^\mu_{\alpha\beta} 
g_{l,b} (\vec k)\psi_\beta(k)}.
\eea
It is more convenient to represent each column of the matrix form 
$n^{\mu,b}$ ($b=1,2$) as a 3-vector in spin space as
\bea
\vec n_{1}=n^{\mu, 1}, \ \ \ \vec n_{2}=n^{\mu, 2}.
\eea
For $l=1$, $\vec n_{1,2}$ are just the spin currents along the
$x,y$ directions respectively.
When $l\ge 2$, $\vec n_{1,2}$ denote the spin multipole components
at the level $l$ on the Fermi surface.
$\vec n_1 \pm i\vec n_2$ carry the orbital angular momenta $L_z=\pm l$
respectively.
In other words, $\vec n_{1,2}$ are the counterpart of the spin-moment
in the $l$-th partial wave channel in momentum space.

The mean field Hamiltonian, {\it i.e.\/} for a state with a uniform 
order parameter, can be decoupled as
\bea
H_{MF}&=&\int \frac{d^2 \vec k}{(2\pi)^2}
\psi^\dagger_\alpha (\vec k) \Big\{\epsilon(\vec k)-\mu
-\Big( \vec n_1 \cos( l\theta_k)   \nonumber \\
&+&   \vec n_2 \sin (l\theta_k)\Big)
\cdot \vec \sigma \Big\} \psi_\beta(\vec k)
+\frac{|n_1|^2 +|n_2|^2}{2|f^a_l|}.\nonumber \\
&&
\label{eq:meanfield}
\eea
This mean field theory is valid when the interaction range
$\xi\approx \sqrt{\kappa  |f^a_l|}$
is much larger than the inter-particle distance $d \approx 1/k_F$.  
The actual validity of mean field theory at quantum criticality requires 
an analysis of the effects of quantum fluctuations which are not included 
in mean field theory~\cite{hertz1976,millis1993}. 
In this theory, just as the case of Ref.~\cite{oganesyan2001}, 
the dynamic critical exponent turns out to be $z=3$ and mean field 
theory appears to hold even at quantum criticality.

Taking into account that  $|f_1^a|\sim 1/N(0)$ around the transition
point, we introduce a dimensionless parameter $\lambda$ to denote
the above  criterion as
\bea
\lambda=\frac{\kappa k_F^2}{ N(0)} \gg 1.
\eea
Finally, notice that, in the $p$-wave channel, the Hamiltonian Eq. \eqref{eq:meanfield} 
can be formally represented through an $SU(2)$ non-Abelian gauge field  minimally-coupled to the fermions
\bea
H_{mf}&=&\int d^2 \vec r~ \frac{1}{2m}  \psi^\dagger (\vec r) 
(-i \hbar \nabla^a - m  A_a^\mu (\vec r) \sigma^\mu)^2 \psi(\vec r)\nn \\
&-& \frac{m}{2} 
 \psi^\dagger (\vec r) \psi(\vec r)
A^\mu_a(\vec r) A^\mu_a (\vec r).
\label{eq:gaugefm}
\eea
where the gauge field is defined as
\bea
A^\mu_a (\vec r) \sigma^\mu = n^{\mu a} (\vec r) \sigma^\mu.
\eea
Notice, however, that this is an approximate effective local gauge
invariance which only holds for a theory with a
{\em linear dispersion relation}, 
and that it is manifestly broken by non-linear corrections, 
such as the quadratic term of Eq. \eqref{eq:gaugefm}, and the cubic 
terms included in the dispersion $\epsilon(\vec k)$ of 
Eq. \eqref{eq:dispersion}.

\section{Ginzburg-Landau free energy}
\label{sec:GLfree}

\subsection{The 2D systems}
In order to analyze the possible ground state configuration 
discussed in Section \ref{sec:intro},
we construct the G-L free energy in 2D in the arbitrary $l$-wave channel.
The symmetry constraint to the G-L free energy is as follows.
Under time-reversal (TR) and parity transformations,
$\vec n_{1,2}$ transform, respectively,  as
\bea
T \vec n_{1,2} T^{-1}= (-)^{l+1} \vec n_{1,2}, \ \ \ P \vec n_{1,2} P^{-1}=
(-)^l \vec n_{1,2}.
\eea
Under the $SO_S(3)$ rotation $R_{\mu\nu}$ in the spin channel,
$\vec n_{1,2}$ transform as
\bea
n_{\mu,1}\rightarrow R_{\mu\nu} n_{\nu,1}, \ \ \,
n_{\mu,2}\rightarrow R_{\mu\nu} n_{\nu,2},
\eea
On the other hand, under a uniform rotation by an angle $\theta$ about the 
$z$-axis, in the orbital channel, the order parameters $\vec n_a$ transform as
\bea
\vec n_1 &\rightarrow&  ~~ \cos (l\theta)~ \vec n_1+ \sin ( l \theta)~ \vec n_2,
\nonumber \\
\vec n_2 &\rightarrow&  -\sin (l\theta)~ \vec n_1+ \cos (l \theta)~ \vec n_2. 
\eea
Thus, the order parameter fields $\vec n_1$ and $\vec n_2$ are invariant 
under spatial rotations by $2\pi/l$, and change sign under a rotation 
by $\pi/l$. In the $\alpha$-phase this change change be compensated
by flipping the spins.

In order to maintain the $SO_L(2)\otimes SO_S(3)$ symmetry, up to quartic 
terms in the order parameter fields $\vec n_a$, 
the uniform part of the GL free energy has the form
\bea
F(n)&=& r \tr (n^T n) + (v_1+\frac{v_2}{2})  [\tr (n^T n)]^2
-\frac{v_2}{2} \tr [(n^T n)^2] \nonumber \\
&=& r (|\vec n_1|^2 +|\vec n_2|^2 ) 
+ v_1 (|\vec n_1|^2 +|\vec n_2|^2 )^2   \nonumber \\
&+& v_2 |\vec n_1 \times \vec n_2|^2.
\label{eq:free2D}
\eea
The coefficients $r,v_1,v_2$ will be presented in Eq. \eqref{eq:2Dcoeff}
by evaluating the ground state energy of the mean field Hamiltonians
Eq. \eqref{eq:Halpha} and Eq. \eqref{eq:Hbeta}. 
The Pomeranchuk instability occurs at $r<0$, {\it i.e.\/}, $F_{l}^a<-2 ~(l\ge 1)$.
Furthermore, for  $v_2>0$, the ground state is the $\alpha$-phase
which favors $\vec n_1 \parallel \vec n_2$, while
leaving the ratio of $|\vec n_1|/|\vec n_2|$ arbitrary.
On the other hand, for $v_2<0$ we find a $\beta$-phase, which favors 
$\vec n_1 \perp \vec n_2$ and $|\vec n_1|=|\vec n_2|$.

The gradient terms are more subtle. 
We present the gradient terms of the GL free energy for $l=1$ as follows
\bea
F_{grad}(n)&=&\gamma_1 \tr [\partial_a n^T \partial_a n]
+\gamma_2 \epsilon_{\mu\nu\lambda} n^{\mu a}n^{\nu b} \partial_a
n^{\lambda b} \nn \\
&=&\gamma_1 (\partial_a \vec n_b \cdot \partial_a \vec n_b )
+\gamma_2 \Big \{ (\partial_x \vec n_2 -\partial_y \vec n_1)\nonumber\\
&\cdot& (\vec n_1 \times \vec n_2) \Big \} 
\label{eq:GL2}.
\eea 
For simplicity, as in Ref.~\cite{oganesyan2001}, we have neglected the difference between two Frank constants
and only present one stiffness coefficient $\gamma_1$. (This approximation is accurate only near the Pomeranchuk quantum critical point.)
More importantly, because  $n^{\mu,b}$ is odd under parity
transformation for $l=1$, a new $\gamma_2$ term appears, which is 
of cubic order in the order parameter field  $n^{\mu,b}$, and it is linear in derivatives. This term is allowed by all the symmetry
requirements, including time reversal, parity, and rotation symmetries.
This term has no important effects in the disordered phase and
in the $\alpha$-phase,
but it leads to a Lifshitz-like inhomogeneous
ground state with spontaneous chirality in the $\beta$-phase in which parity 
is spontaneously broken.
We will discuss  this effect in detail in Section \ref{sec:beta}.
The coefficient of $\gamma_{1,2}$ will be presented in Eq. \eqref{eq:gradcoeff}.
Similarly, for all the odd values of $l$, we can 
write a real cubic $\gamma_2$ term satisfying all the symmetry constraints as
\bea
 \gamma_2 \epsilon_{\mu\nu\lambda} n^{\mu a} n^{\nu b}
(i)^l g_a (\hat \nabla) n^{\lambda b}.
\eea
However, this term corresponds to high order corrections,
and is negligible (irrelevant) for $l\ge 3$.

Similarly to the approximate gauge symmetry for the fermions
 in Eq. \eqref{eq:gaugefm},
the $\gamma_2$ term can also be reproduced by a non-Abelian gauge potential
defined as
\bea
i A^\lambda_a (x) (T^\lambda)_{\mu\nu}=\epsilon_{\lambda\mu\nu} n^{\lambda a}
(x),
\eea
where $T^\lambda_{\mu\nu}=-i\epsilon_{\lambda\mu\nu}$ is the generator
of the $SU(2)$ gauge group in the vector representation.
Then Eq. \eqref{eq:GL2} can be written as
\bea
F_{grad}(n)&=&\gamma_1 \Big\{(\partial_a \delta_{\mu\nu}
- i g A^\lambda_a (T^\lambda)_{\mu\nu})~ n^{\nu b}\Big\}^2\nn \\
&+& \gamma_1 g^2 (\epsilon_{\lambda\mu\nu} n^{\lambda a} n^{\nu b})^2
\nn \\
&=&\sum_{ab}\Big\{ \gamma_1   (\partial_a \vec n_b +
 g ~\vec n_a \times \vec n_b)^2 \nn \\
&-&\gamma_1 g^2 |\vec n_a\times  \vec n_b|^2 \Big\},
\label{eq:gaugeGL}
\eea
with $g=\gamma_2/(2\gamma_1)$.

\subsection{The 3D systems}
In 3D, the order parameter in the $F_l^a$ channel Pomeranchuk instabilities
can be similarly represented by a $3\times (2l+1)$ matrix.
Here we only consider the simplest case of the $p$-wave channel 
instability $(l=1)$, which has been studied in Ref. 
\cite{hirsch1990,oganesyan2001,varma2006,wu2004} under different
contexts.
In the $F_1^a$ channel, the order parameter 
$n^{\mu,i}$ is a $3\times 3$ real matrix defined as
\bea
n^{\mu,b}=|f^a_1|\int \frac{d^3 \vec k}{(2\pi)^3}
\avg{\psi^\dagger_\alpha(k) 
\sigma^\mu_{\alpha\beta}  k^b  \psi_\beta(k)}.
\eea
The difference between $n^{\mu,i}$ and the triplet $p$-wave pairing
order parameter in the $^3$He system \cite{leggett1975,vollhardt1990}
is that the former is defined in the particle-hole channel, and thus is real.
By contrast, the latter one is defined in the particle-particle channel
and is complex.
Each column of the matrix form $n^{\mu,b} (b=x,y,z)$ can be viewed
as a $3$-vectors in the spin space as
\bea
\vec n_1 = n^{\mu,1}, \ \ \ \vec n_2 = n^{\mu,2}, \ \ \
\vec n_3 = n^{\mu,3},
\eea
which represents the spin current in the $x,y$ and $z$ directions
respectively.
The  G-L free energy in Ref. \cite{wu2004} can be reorganized as
\bea
F(n)&=&  r^\prime \tr (n^T n) + (v_1^\prime+\frac{v^\prime_2}{2}) 
 [\tr (n^T n)]^2
-\frac{v_2^\prime}{2} \tr [(n^T n)^2] \nonumber \\
&=&r^\prime (|\vec n_1|^2 +|\vec n_2|^2 + |\vec n_3|^2 ) 
+ v^\prime_1 (|\vec n_1|^2 \nonumber \\
&+&|\vec n_2|^2 +|\vec n_3|^2)^2   
+ v_2^\prime \Big\{ |\vec n_1 \times \vec n_2|^2 \nonumber \\
&+& |\vec n_2 \times \vec n_3|^2+
|\vec n_3 \times \vec n_1|^2 \Big\},
\label{eq:free3D}
\eea
where the coefficients $r^\prime, v^\prime_{1,2}$ will be presented
in Eq. \eqref{eq:3Dcoeff}.
Similarly, the $\alpha$-phase appears at $v_2>0$ which favors that
$\vec n_1 \parallel \vec n_2 \parallel \vec n_3$, and leaves their
ratios arbitrary.
The $\beta$-phase appears at $v_2<0$ which favors that
vectors $\vec n_{1,2,3}$ are perpendicular to each other
with equal amplitudes $|\vec n_1|=|\vec n_2|=|\vec n_3|$. 

Similarly, we present the gradient terms in the G-L free energy
as
\bea
F_{grad}(n)&=&\gamma^\prime_1 \tr[\partial_a n^T \partial_a n]
+\gamma^\prime_2 \epsilon_{\mu\nu\lambda} n^{\mu i} n^{\nu j} 
\partial_j n^{\lambda i}, \nonumber \\
\label{eq:3Dinstab}
\eea
where the coefficient $\gamma^\prime_{1,2}$ will be presented at
Eq. \eqref{eq:gamma3d}. 
Again, we neglect the difference among three Frank constants.
The $\gamma^\prime_2$ term can be represented as
\bea
\gamma_2^\prime \Big\{\!\!\!& &(\partial_x \vec n_2 -\partial_y \vec n_1) 
\cdot (\vec n_1 \times \vec n_2)\nonumber \\
&&+
(\partial_y \vec n_3 -\partial_z \vec n_2)  
\cdot (\vec n_2 \times \vec n_3)\nonumber \\
&&+
(\partial_z \vec n_1 -\partial_x \vec n_3) 
\cdot (\vec n_3 \times \vec n_1) \Big\}.
\eea
It can also be represented in terms of the non-Abelian gauge
potential as in Eq. \eqref{eq:gaugeGL}.

\section{Mean field phases in the $F^a_l$ channel}
\label{sec:meanfield}

In this section, we discuss the solution to the mean field Hamiltonian
Eq. \eqref{eq:meanfield}, for the ordered $\alpha$ and $\beta$-phases 
in both 2D and 3D.

\subsection{The 2D $\alpha$-phases}
\label{subsec:alpha}

\begin{figure}
\psfrag{l=1}{$l=1$}
\psfrag{l=2}{$l=2$}
\psfrag{s}{$\vec s$}
\psfrag{a-phase}{\small $\alpha$-phase}
\psfrag{dk1}{$\delta k_\uparrow$}
\psfrag{dk2}{$\delta k_\downarrow$}
\begin{center}
\includegraphics[width=0.45\textwidth]{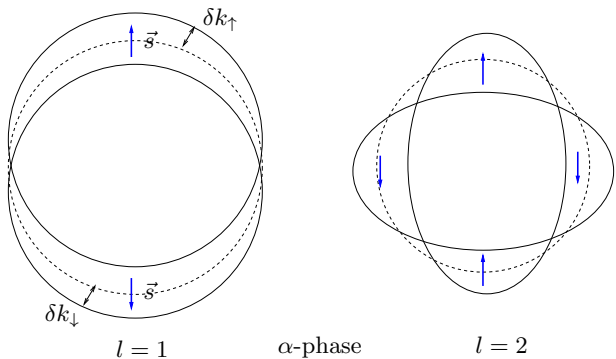}
\end{center}
\caption{The $\alpha$-phases in the $F^a_1$ and $F^a_2$ channels. 
The Fermi surfaces exhibit the $p$ and $d$-wave distortions, respectively.
}\label{fig:alpha}
\end{figure}  

The $\alpha$-phase is characterized by an anisotropic distortion of the Fermi surfaces of up and down spins. In this phase $s_z$ is as a good quantum number.
It is a straightforward generalization of the nematic Fermi liquid 
for the case of the spin channel \cite{oganesyan2001}.
For example, the Fermi surface structures at $l=1,2$ are depicted 
in Fig. \ref{fig:alpha}. The quadrupolar, $l=2$, case is the nematic-spin-nematic phase~\cite{oganesyan2001,kivelson2003}.
Without loss of generality, we choose $\vec n_1= \bar n\; \hat z$, and
$|n_2|=0$.
The mean field Hamiltonian $H_\alpha$ for the $\alpha$-phase becomes
\bea
H_{\alpha,l}&=& \int \frac{d^2 k}{(2\pi)^2}
 \psi^\dagger(\vec k) [\epsilon(\vec k)-\mu
-\bar n \cos (l \theta_k) \sigma_z) ] \psi(\vec k).
\nonumber \\
\label{eq:Halpha}
\eea
The dispersion relations for the spin up and down electrons become
\bea
\xi_{\uparrow,\downarrow}(\vec k)&=&\epsilon(k)-\mu  \mp
\bar n \cos (l \theta_k),
\eea
The value of $\bar n$ can be obtained by solving the
self-consistent equation in the $\alpha$-phase
\bea
\frac{\bar n}{|f^a_l(0)|}&=&\int \frac{d \vec k}{(2\pi)^2}
\{n_f(\xi_\uparrow(\vec k))-n_f(\xi_\downarrow(\vec k))\} \cos (l \theta_k).
\nn \\
\label{eq:sfconalpha}
\eea
where $n_f(\xi_{\uparrow}(k))$ and $n_f(\xi_\downarrow(k))$ are the Fermi functions for the up and down electrons respectively.
The distortions of the Fermi surfaces of the spin up and spin down bands 
are given by an angle-dependent part of their Fermi wave vectors:
\bea
\frac{\delta k_{F\uparrow,\downarrow}(\theta)}{k_F} 
&=& \frac{2 a-1}{4} x^2  \pm (x+\frac{a-2 a^2}{2} x^3 ) \cos l \theta
 \nonumber \\
&-& a x^2  \cos^2 l \theta \pm (2 a^2 -b ) x^3 \cos^3 l \theta +O(x^4),  \nonumber \\
\label{eq:fsdistalpha}
\eea
where we introduced the dimensionless parameter $x=\bar n/(v_F k_F)$, where $v_F$ and $k_F$ are the Fermi velocity and the magnitude of the Fermi wave vector in the FL phase. This solution holds for small distortions and it is accurate only close to the quantum phase transition. By inspection we see that in the $\alpha$-phases the total spin polarization vanishes. More importantly, under a rotation by $\pi/l$, the charge and spin components of the order parameter both change sign, {\it i.e.\/} a rotation by $\pi/l$ is equivalent to a reversal of the spin polarization.

The single particle fermion Green function in the $\alpha$-phase at wave vector $\vec k$ and Matsubara frequency $\omega_n$ is
\bea
G(\vec k, i\omega_n)=\frac{1}{2} \Big\{
\frac{1+\sigma_z} {i\omega_n-\xi_\uparrow(\vec k)}
+\frac{1-\sigma_z} {i\omega_n-\xi_\downarrow(\vec k)}
\Big\}.
\eea

\subsection{The 2D $\beta$-phases}
\begin{figure}
\psfrag{s}{$\vec s$}
\psfrag{dk1}{$\delta k_\uparrow$}
\psfrag{dk2}{$\delta k_\downarrow$}
\begin{center}
\subfigure[~Rashba ($w=1$)]{\includegraphics[width=0.3\textwidth]{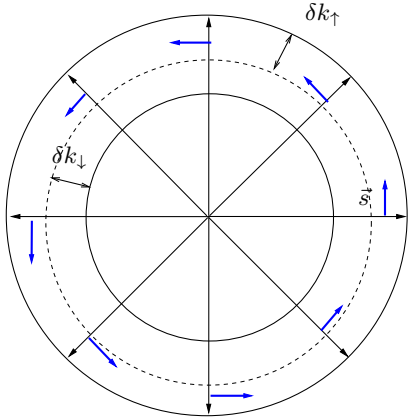}}

\subfigure[~Dresselhaus ($w=-1$)]{\includegraphics[width=0.3\textwidth]{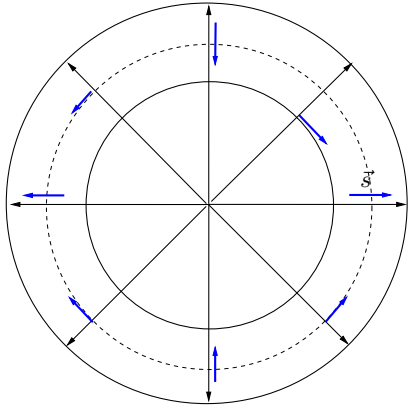}}
\end{center}
\caption{The $\beta$-phases in the $F^a_1$ channel. Spin configurations
exhibit the vortex structures in the momentum space
with winding number $w=\pm 1$, which correspond to 
Rashba and Dresselhaus SO coupling respectively.
}\label{fig:vortex1}
\end{figure}   

\begin{figure}
\psfrag{s}{$\vec s$}
\psfrag{dk1}{$\delta k_\uparrow$}
\psfrag{dk2}{$\delta k_\downarrow$}
\begin{center}
\subfigure[~$w=2$]{\includegraphics[width=0.43\linewidth]{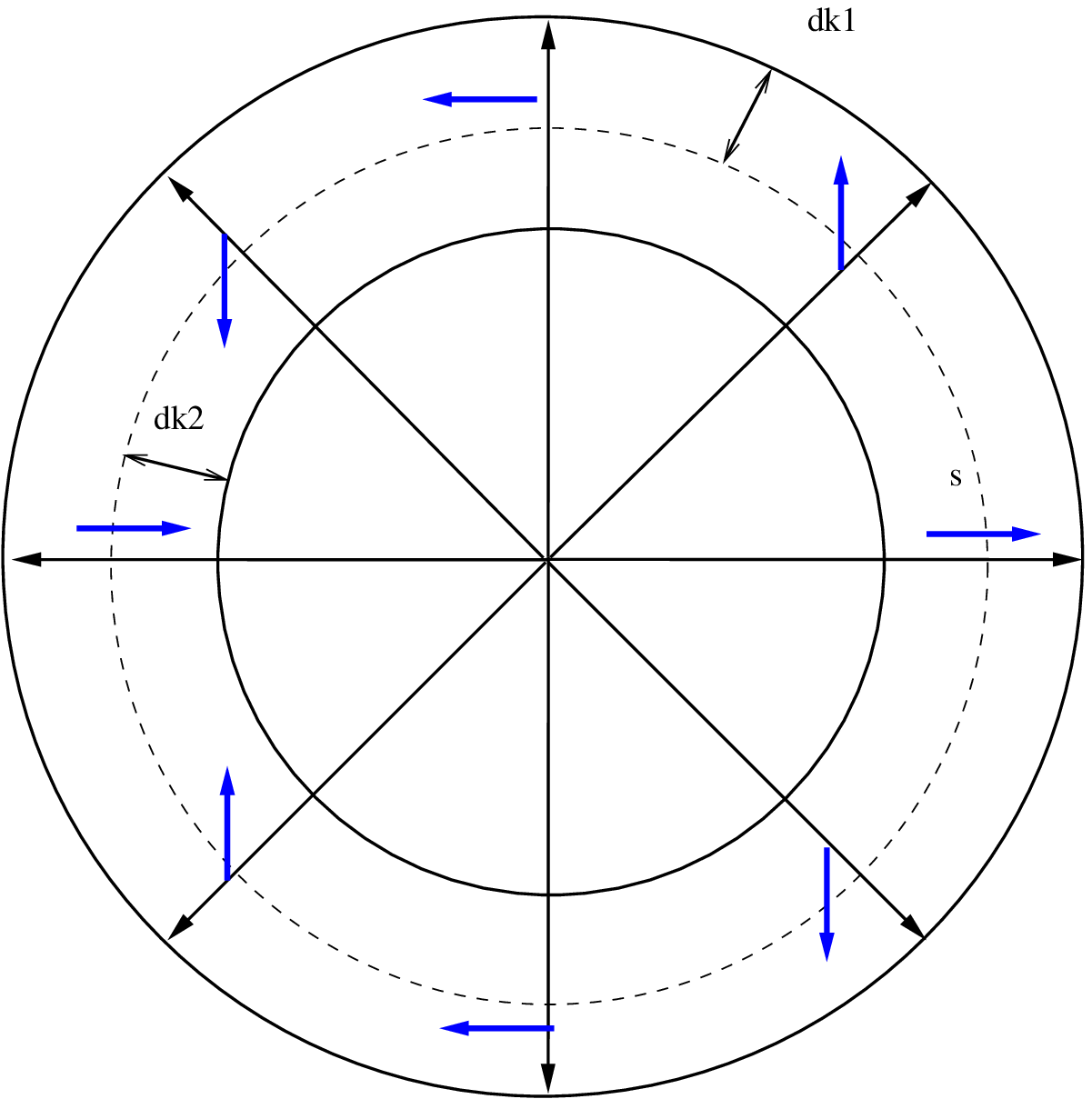}}
\subfigure[~$w=-2$]{\includegraphics[width=0.43\linewidth]{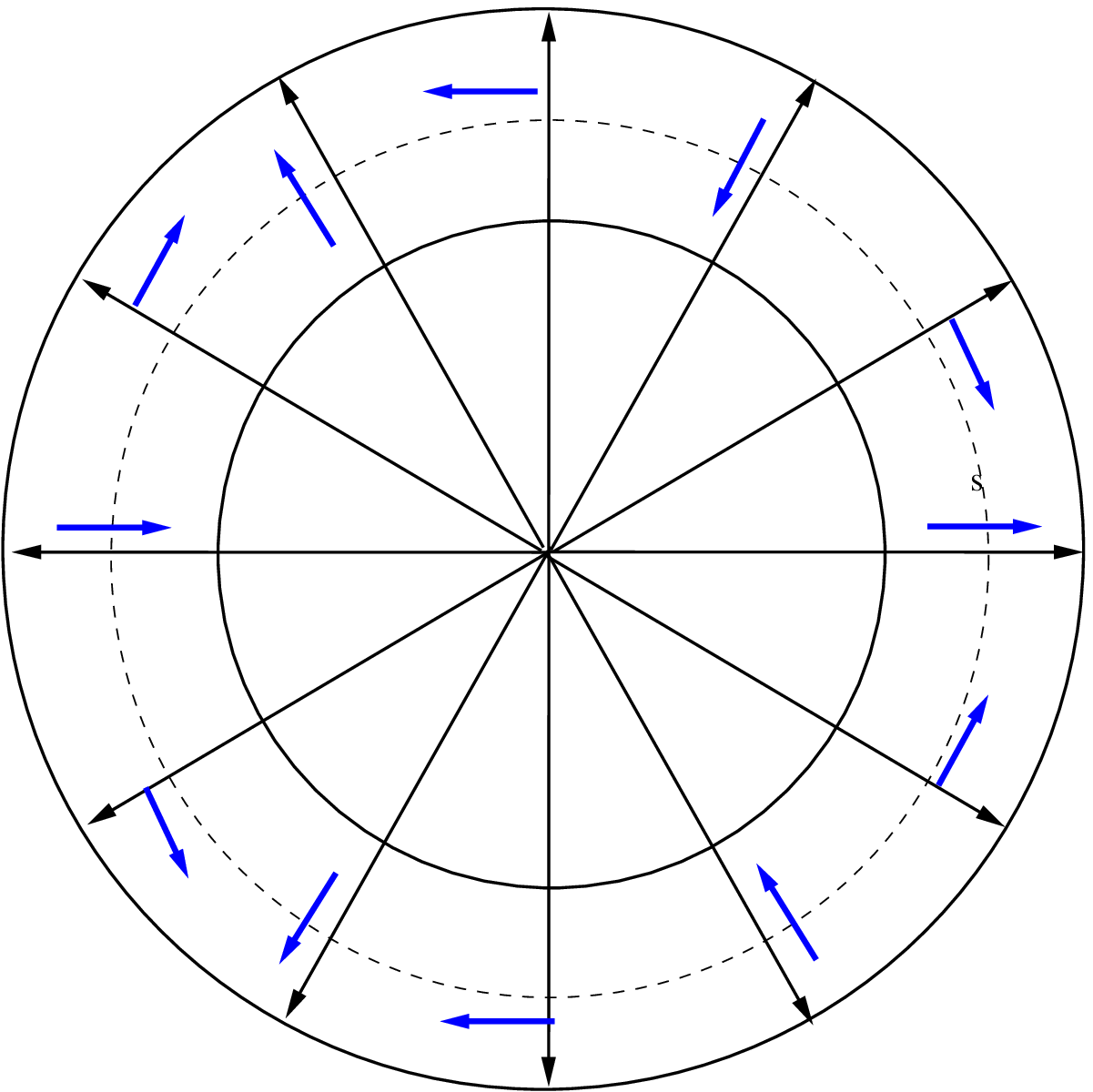}}
\end{center}
\caption{The $\beta$-phases in the $F^a_2$ channel. Spin configurations
exhibit the vortex structure with winding number $w=\pm 2$.
These two configurations can be transformed to each other
by performing a rotation around the $x$-axis with the angle of $\pi$.
}\label{fig:vortex2}
\end{figure}

\begin{figure}
\psfrag{S2}{$S^2$}
\psfrag{n1}{$\vec n_1$}
\psfrag{n2}{$\vec n_2$}
\begin{center}
\includegraphics[width=0.5\linewidth]{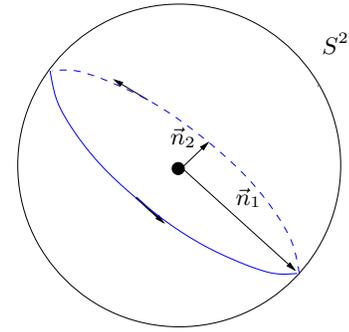}
\end{center}
\caption{The spin configurations on both Fermi surfaces in the
$\beta$-phase ($F^a_l$ channel) map to a large circle on an 
$S^2$ sphere with the winding number $l$.
}\label{fig:vortex3}
\end{figure}

The  $\beta$-phase appears for $v_2<0$, which
favors $\vec n_1 \perp \vec n_2$ and $|n_1|=|n_2|$.
Like the case of ferromagnetism, the Fermi surfaces split into 
two parts with different volumes, while each one still keeps 
the round shape undistorted.
However, an important difference exists between the $\beta$-phase
at $l\ge 1$ and the ferromagnetic phase.
In the ferromagnet, the spin is polarized along a fixed uniform direction, which gives rise to a net spin moment.
On the other hand, in the $\beta$-phase with orbital angular momentum $l\ge 1$, {\em the
spin winds around the Fermi surface} exhibiting a vortex-like structure
in momentum space. Consequently in the $\beta$-phase the net spin moment is zero, just as it is in the $\alpha$-phase. (Naturally, partially polarized versions of the $\alpha$ and $\beta$-phases are possible but will not be discussed here.)
In other words, it is a spin nematic, high partial wave channel 
generalization of ferromagnetism.
In the previously studied cases of the $F^a_1$ channel in Ref.
\cite{wu2004}, it was shown that in this phase effective Rashba and Dresselhaus terms are dynamically generated in single-particle Hamiltonians.
The ground state spin configuration exhibits, in momentum space, a vortex structure 
with winding number $w=\pm1$  depicted in Fig. \ref{fig:vortex1}.

Here we generalize the vortex picture in momentum space
in the $F^a_1$ channel to a general $F^a_l$ channel.
We assume $|n_1|=|n_2|=\bar n$.
Without loss of generality, we can always perform an $SO(3)$ rotation
in spin space to set $\vec n_1 \parallel \hat x$,
and $\vec n_2 \parallel  \hat y$.
Then, much as in the B phase of $^3$He, the mean field Hamiltonian $H_{\beta,l}$ for the $\beta$-phase in angular momentum channel $l$
can be expressed through a $d$-vector, defined by
\bea
\vec d(\vec k) =\left(\cos (l \theta_k) , \sin (l\theta_k) , 0\right),
\label{eq:dvector}
\eea
as follows
\bea
H_{\beta,l}&=& \int \frac{d^2 k}{(2\pi)^2}
\psi^\dagger(\vec k) \left[\epsilon(\vec k)-\mu
-\bar n \vec d (\theta_k) \cdot \vec \sigma) \right] \psi(\vec k),
\nonumber \\
\label{eq:Hbeta}
\eea
where $\vec d (\theta_k)$ is the spin quantization axis for
single particle state at $\vec k$.
The saddle point value of $\bar n$ can be obtained by solving the
self-consistent equation 
\bea
\frac{2 \bar n}{|f^a_l(0)|}&=&\int \frac{d^2 \vec k}{(2\pi)^2}
\left(n_f(\xi_\uparrow(\vec k))-n_f(\xi_\downarrow(\vec k)\right),
\eea
where the single particle spectra read
\bea
\xi_{\uparrow,\downarrow}(\vec k)=\epsilon(\vec k)-\mu\pm\bar n,
\eea
which is clearly invariant under spacial rotations.
The Fermi surface splits into two parts with 
\bea
\frac{\delta k_{F,\uparrow,\downarrow}}{k_F}
=\pm x -\frac{x^2}{2} \pm (a-b) x^3 + O(x^4).
\eea
The single particle Green function in the $\beta$-phase reads
\bea
G_{ l}(\vec k, i\omega_n)=\frac{1}{2} \Big\{
\frac{1+\vec \sigma \cdot \hat d} {i\omega_n-\xi_\uparrow(\vec k)}
+\frac{1-\vec \sigma \cdot \hat d} {i\omega_n-\xi_\downarrow(\vec k)}
\Big\},
\eea 
where $\vec \sigma \cdot \vec d (\hat k)$ is 
the $l$-th order  helicity operator.
Each Fermi surface is characterized by the eigenvalues $\pm 1$ of the helicity 
operators  $\vec \sigma \cdot \vec d (\hat k)$.

From the mean field theory of $\alpha$ and $\beta$-phases, we
can calculate the coefficients of the G-L theory, 
Eq.\eqref{eq:free2D}, as
\bea
r&=&\frac{N(0)}{2 }(\frac{1}{|F^a_l|}-\frac{1}{2}), \nonumber\\
v_1&=& \frac{N(0)}{32} \Big\{[\frac{N^\prime(0)}{N(0)}]^2
-\frac{N^{\prime\prime}(0)}{2N(0)} \Big\} \nonumber \\
&=&(1-a -2a^2 +3 b)\frac{N(0)}{32 v_F^2 k_F^2},~ \nonumber \\
v_2&=& \frac{N^{\prime\prime}(0)}{48}
=(-a+2 a^2-b) \frac{N(0)}{8 v_F^2 k_F^2},
\label{eq:2Dcoeff}
\eea
where $v_{1,2}$  do not depend on the value of $l$ 
at the mean field level.
$N^\prime (0)$ and $N^{\prime\prime}(0)$ are the first and second
order derivatives of density of states at the Fermi energy
$E_f$ respectively. 
They are defined as
\bea
N^\prime(0)&=& \frac{d N}{d \epsilon} |_{\epsilon=E_f}
=(1-2a) \frac{N(0)}{v_f k_f}
, \nonumber \\
N^{\prime\prime}(0)&=& \frac{d^2 N}{d \epsilon^2} |_{\epsilon=E_f}
=6(-a+2a^2-b) \frac{N(0)}{v_f^2 k_f^2}. \ \ \ \ \ \
\eea
Both of them only depend on the non-linear dispersion
relation up to the cubic order as kept 
in  Eq. \ref{eq:dispersion}.

It is worth to stress that the coefficients of Eq. \eqref{eq:2Dcoeff}
were calculated (within this mean field theory) at {\em fixed density}. 
Similar coefficients  were obtained in the spinless system analyzed
in Ref. \cite{oganesyan2001} at fixed chemical potential. 
The is a subtle difference between these two settings in the behavior 
of the quartic terms. At fixed chemical potential the sign of $b$,
the coefficient of the cubic term in the free fermion dispersion relation, 
is crucial for the nematic phase to be stable. 
However, as can be seen in Eq.\eqref{eq:2Dcoeff}, the sign of the 
coefficient of the quartic term $v_1$ is determined 
by several effects: the coefficients $a$ and $b$, and 
that of an extra 
(additive) contribution which originates from the curvature  of the
Fermi surface and hence scales as $N(0)/k_F^2$.
It has been noted 
in Refs. \cite{Kee2003a,khakvine04,Kee2006} that  the nematic
 instability for lattice systems may be a continuous quantum phase 
transition or a first order transition, in which case it involves
 a change in the topology of the Fermi surface. As shown above, 
this dichotomy is the result of the interplay of the single particle 
dispersion and effects due to the curvature of the Fermi surface. 
The same considerations apply to the coefficients that we will
 present in the following subsection.

The ground state spin configuration in the $\beta$-phase  exhibits a vortex structure 
with winding number $w=l$ in momentum space. The case of $w=2$ is 
depicted in Fig. \ref{fig:vortex2} A.
Interestingly, in the case of $l=3$,
after setting $\vec d=(\cos(3\theta+\pi/2),\sin(3\theta+\pi/2),0)$,
the effective single particle Hamiltonian becomes
\be
H_{MF}(k)=\epsilon(k)+\bar n \left[-\sin (3\theta_k) \sigma_x +\cos (3\theta_k )
\sigma_y \right].
\ee
This single particle Hamiltonian has the same form as is that of the heavy hole band of the 2-dimensional
$n$-doped GaAs system \cite{winkler2000,schliemann2005}, which is results from SO coupling.

Now we discuss the general configuration of the $d$-vector in the 
$\beta$-phase in the $F^a_l$ channel.
$\vec n_1$ and $\vec n_2$ can be any two orthogonal unit
vectors on the $S^2$ sphere. 
The plane spanned by $\vec n_{1,2}$ intersects the $S^2$ sphere at any
large circle as depicted in Fig. \ref{fig:vortex3}, 
which can always be obtained by performing
a suitable $SO(3)$ rotation from the large circle in the $xy$ plane.
The spin configuration around the Fermi surface maps to this
large circle with the winding number of $l$.
Furthermore, the configuration of winding number $\pm l$ are equivalent
to each other up to rotation of $\pi$ around a diameter of the
large circle.
For example, the case of $w=-2$ is depicted in Fig. \ref{fig:vortex2} B,
which can be obtained from that of $w=2$ by performing such a rotation
around the $\hat x$-axis.
Similarly, with the $SO_S(3)$ symmetry in the spin space, the configurations
with $w=\pm l$ are topologically equivalent to each other.
However, if the $SO_S(3)$ symmetry is reduced to $SO_S(2)$ because of the existence of an explicit
easy plane magnetic anisotropy (which is an effect of SO interactions at the single particle level), or by an external magnetic field $\vec B$,
then the two configurations with $w=\pm l$ belong to two distinct
topological sectors.

\subsection{The 3D instabilities of the $p$-wave spin triplet channel}

The mean field theory for the Pomeranchuk instability in the $F^a_1$
channel has been studied in Ref. \cite{wu2004}.
To make the paper self-contained, here we summarize the main results.

In the  $\alpha$-phase, taking the special case $n^{\mu a}=\bar n
\delta_{\mu z} \delta_{a z}$, the mean field Hamiltonian reads
\bea
H_{2D, \alpha}&=&\int \frac{d^3 k}{(2\pi)^3}
 \psi^\dagger(\vec k) \left[\epsilon(\vec k)-\mu
-\bar n \sigma_z \cos \theta_k ) \right] \psi(\vec k),
\nonumber \\
\label{eq:Halpha3D}
\eea
where $\theta$ is the angle between $\vec k$ and $z$-axis.
The Fermi surfaces for the two spin components are distorted in an
opposite way as 
\bea
\frac{\Delta k_{F\uparrow,\downarrow}(\theta)}{k_F}&=&
\frac{1}{3} (1-a) x^2 \pm  [x+\frac{2}{3} a (1-a)^3 x^3] \cos \theta\nonumber \\
&-&a x^2 \cos^2 \theta \pm (2 a^2-b) x^3 \cos^3 \theta +O(x^4)
\nonumber \\
\eea

In the $\beta$-phase,  rotational symmetry is preserved and a SO interaction is
dynamically generated.
With the ansatz $n^{\mu a}=\bar n \delta_{\mu a}$,
the MF Hamiltonian reduces to
\bea H_{3D, \beta}=
\sum_k \psi^\dagger(k) (\epsilon(k)-\mu -
\bar n \vec \sigma \cdot \hat k
) \psi(k). \label{BMF}
\eea
The single particle states can be classified according to the
eigenvalues $\pm 1$ of the helicity
operator $\vec \sigma\cdot \hat k$, with dispersion relations
$\xi^B(k)_{\uparrow,\downarrow}=\epsilon(k)-\mu\pm \bar n$. 
The Fermi surfaces split into two parts, but still keep
the round shape for two helicity bands with
\be
\frac{\Delta k_{F\uparrow,\downarrow}}{k_F}= 
\pm x - x^2 \pm (2a-b) x^3 +O(x^4).
\ee
The $\beta$-phase is essentially isotropic.
The orbital angular momentum $\vec  L$ and spin $\vec S$
are no longer separately conserved, but the total
angular momentum $\vec J= \vec L+\vec S=0$ remains conserved
instead.
For the general case of $n_{\mu a}=\bar n D_{\mu a}$, it is equivalent to
a redefinition of spin operators as $S^\prime_\mu=S_\nu D_{\nu a}
\delta_{a\mu}$, thus Fermi surface distortions remain isotropic
and $\vec J^\prime=\vec L +\vec S^\prime$ is conserved.

From the above mean field theory, we can calculate  the
coefficients in Eq. \eqref{eq:free3D} as 
\bea
r^\prime&=&\frac{N(0)}{2 }(\frac{1}{|F_1^a|}-\frac{1}{3}), 
\nonumber \\
v^\prime_1&=&\frac{N(0)}{24} \Big\{\frac{1}{3} [\frac{N^\prime(0)}{N(0)}]^2
-\frac{N^{\prime\prime}(0)}{5N(0)} \Big\}\nonumber \\
&=&
\frac{N(0)}{180 v_F^2 k_F^2}
(7-2a-8a^2+9b),
\nonumber \\
v^\prime_2&=&\frac{N_2(0)}{90}=
\frac{N(0)}{45 v_F^2 k_F^2}(1-6a+6a^2-3b), \nonumber \\
\label{eq:3Dcoeff}
\eea
where 
\bea
N^\prime(0) 
&=&\frac{ d N}{d \epsilon}|_{\epsilon=E_f}=\frac{2-2a}{v_f k_f} N(0), 
\nonumber \\
N^{\prime\prime}(0)&=& \frac{ d^2 N}{d \epsilon^2}|_{\epsilon=E_f}
=\frac{2(1-6a+6a^2-3b)}{v^2_f k^2_f}N(0). \nonumber \\
\eea
Once again, the caveats of the previous subsection on the sign of the 
coefficients of the quartic terms apply here too.

\section{Goldstone manifolds and topological defects}
\label{sec:topology}

\begin{figure}
\begin{center}
\includegraphics[width=0.3\textwidth]{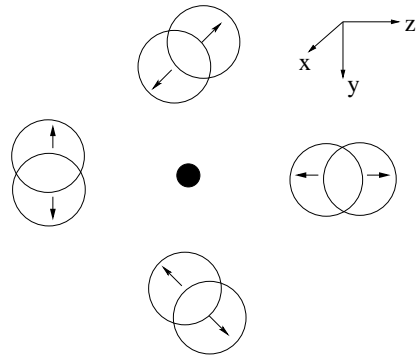}
\end{center}
\caption{The half-quantum vortex with the combined spin-orbit distortion
for the $\alpha$-phase at $l=1$. The triad denotes the direction
in the spin space.
}\label{fig:hqvalpha}
\end{figure} 

In this Section we will discuss the topology of the broken symmetry $\alpha$
and $\beta$-phases, and their associated Goldstone manifolds in 2D and 3D.
We also discuss  and classify their topological defects. 
We should warn the reader that the analysis we present here is based only on the static properties of the broken symmetry phases and ignores potentially important physical effects due to the fact that in these systems the fermions remain gapless (although quite anomalous). In contrast, in anisotropic superconductors the fermion spectrum is gapped (up to possibly a set of measure zero of nodal points of the Fermi surface). The physics of these effects will not be discussed here.

\subsection{Topology of the $\alpha$-phases}
\subsubsection{$2D$ $\alpha$-phases}

For the 2D $\alpha$-phases, for which we can set $\avg{ n^{\mu a}}=\bar n 
\delta_{\mu z} \delta_{a 1}$, the system is invariant under
$SO_{S}(2)$ rotation in the spin channel, the $\mathbb{Z}_l$ rotation
with the angle of $2\pi/l$ in the orbital channel, 
and a $\mathbb{Z}_2$  rotation with the angle of $\pi$
around the $x$-axis in the spin channel combined with
an orbital rotation at the angle of $\pi/l$.
Thus the Goldstone manifold is
\bea
&&[SO_L(2)\otimes SO_S(3)]/ [SO_S(2) \ltimes \mathbb{Z}_2 \otimes \mathbb{Z}_l]\nn \\
&=&[SO_L(2)/\mathbb{Z}_l]\otimes [S^2/\mathbb{Z}_2],
\label{eq:gs2dalpha}
\eea
giving rise to  three Goldstone modes:
one describes the oscillation of the Fermi surface, 
the other two describe the spin precession.
Their dispersion relation will be calculated  in Section \ref{subsec:2dalpha}.

Due to the $\mathbb{Z}_2$ structure of the combined spin-orbit rotation,
the vortices in the 2D-$\alpha$-phase can be divided into two classes.
The first class is the $1/l$-vortices purely in the orbital channel
without distortions in the spin channel.
This class of vortices have the same structure as that in the 
Pomeranchuk instabilities in the density channel dubbed the integer
quantum vortex.
On the other hand, another class of vortices as combined spin-orbital
defects exist.
This class of vortices bears a similar structure to that of  the half-quantum
vortex (HQV) in a superfluid with internal spin degrees of freedom
\cite{mcgraw1994,zhou2003,wu2005}.
An example vortex at $l=1$ of this class is depicted 
in Fig. \ref{fig:hqvalpha} where the $\pi$-disclination in the 
orbit channel is offset
by the rotation of $\pi$ around the $x$-axis in the spin channel.
To describe the vortex configuration for the case of the $p$ wave channel, we set up a local reference frame in spin space, and assume that the electron spin is either 
parallel or anti-parallel to
the $z$-axis of this frame, at a point $\vec x$ in real space. Let $\phi=0$ be angular polar coordinate of $\vec x$ with respect to the core of the vortex.
As we
trace a path  in real space around the vortex, the frame in spin space rotates so that the 
spin flips its direction from up to down (or vice versa) as we rotate by an angle of $\pi$. (In the $d$-wave channel the vortex involves a rotation by $\pi/2$.)
Such behavior is a condensed matter example of the Alice-string behavior 
in the high energy physics \cite{schwarz1982,bucher1999}.
Another interesting behavior of HQV is that a pair of half-quantum
vortex and anti-vortex can carry spin quantum number.
This is an global example of the Cheshire charge 
in the gauge theory \cite{mcgraw1994, wu2005}. 
The electron can exchange spin with the Cheshire charged HQV pairs when
it passes in between the HQV pairs.

Due to the  $SO(2)_L\times SO_S(2)$ symmetry in the Hamiltonian, the fluctuations in the
orbital channel are less severe than those in the spin channel.
In the ground state, the spin stiffness should be softer than that in
orbital channel.
As a result, an integer-valued vortex should be energetically favorable
to fractionalize into a pair of HQV.
However, at finite temperatures, in the absence of magnetic anisotropy (an effect that ultimately is due to spin orbit effects at the atomic level) the spin channel is disordered
with exponentially decaying correlation functions, and thus without long range order in the spin  channel.
However, the orbital channel still exhibits the Kosterlitz-Thouless
behavior at 2D where the low energy vortex configurations should be of
HQV.

\subsubsection{$3D$ $\alpha$-phases}

Similarly, the Goldstone manifold in the 3D $\alpha$-phase for $l=1$ 
can be written as
\bea
&&[SO_L(3)\otimes SO_S(3)]/[SO_L(2)\otimes SO_S(2)\ltimes \mathbb{Z}_2]\nn \\
&=&[S^2_L \otimes S^2_S]/\mathbb{Z}_2,
\label{eq:hqv3dalpha}
\eea
where again the $\mathbb{Z}_2$ operation is a combined spin-orbit rotation
at the angle of $\pi$ to reverse the spin polarization and orbital 
distortion simultaneously.
This Goldstone mode manifold gives rise to two Goldstone modes in the density channel
responsible for the oscillations of the distorted Fermi  surfaces, and
another  two Goldstone modes for the spin precessions.
The fundamental homotopy group of Eq. \eqref{eq:hqv3dalpha} reads $\pi_1[S^2_L 
\otimes S^2_S]/\mathbb{Z}_2 =\mathbb{Z}_2$. 
This means that the $\pi$-disclination exists as a stable topological
line defect.
On the other hand, for the point defect in 3D space, the second
homotopy group of Eq. \eqref{eq:hqv3dalpha} reads
$\pi_2([S^2_L \otimes S^2_S]/\mathbb{Z}_2)=\mathbb{Z}\otimes \mathbb{Z}$.
This means both the orbit and spin channels can exhibit monopole (or hedgehog)
structures characterized by a pair of winding numbers $(m,n)$.
As a result of the $\mathbb{Z}_2$ symmetry, $(m,n)$ denotes the same
monopoles as that of $(-m,-n)$.

\subsection{Topology of the $\beta$-phases}

In the 2D $\beta$-phase with $l=1$ and $w=1$ in the $xy$-plane, the ground 
state is rotationally invariant, thus $L_z+\sigma_z/2$ is 
still conserved.
Generally speaking, the 2D $\beta$-phases with the momentum
space winding number $w$ is invariant under the combined rotation
generated by $L_z+ w  \sigma_z/2$.
The corresponding Goldstone manifold is
\bea
[SO_L(2)\otimes SO_S(3)]/ SO_{L+S}(2)=SO(3).
\eea
Three Goldstone modes exist as relative spin-orbit rotations
around $x,y$ and $z$-axes on the mean field ground state.
Similarly, for the 3D $\beta$-phase with $l=1$, the Goldstone mode
manifold is 
\bea
[SO_L(3)\otimes SO_S(3)]/SO(3)_{L+S}=SO(3),
\eea
with three branches of Goldstone modes.
The dispersion relation of these Goldstone modes will be
calculated in Section \ref{sec:beta}.
The vortex-like point defect in  2D and the line defect in 3D
are determined by the fundamental homotopy group $\pi_1 (SO(3))=\mathbb{Z}_2$.
On the other hand, the second homotopy group $\pi_2 (SO(3))=0$,
thus, as usual, no topologically stable point defect exists in 3D.

\section{The RPA analysis in the critical region at zero temperature}
\label{sec:critical}

\begin{figure}
\psfrag{q}{$\vec q$}
\psfrag{atrans}{\small A: transverse modes}
\psfrag{blong}{\small B: longitudinal modes}
\begin{center}
\includegraphics[width=0.4\textwidth]{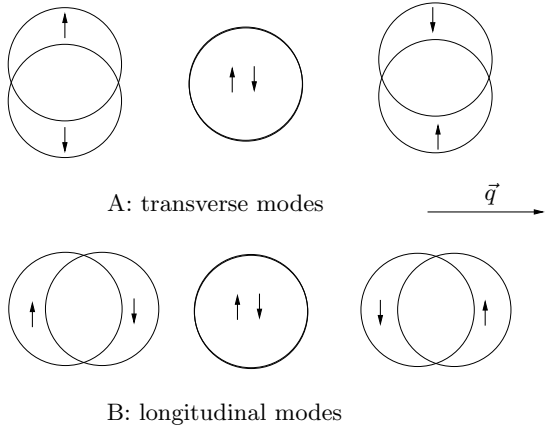}
\end{center}
\caption{The $p$-wave counterpart of paramagnon modes 
in 2D. Taking into account the
spin degrees of freedom, there are 
three (six) transverse triplet modes in  2D (3D), 
and three  longitudinal modes respectively.
}\label{fig:disorder}
\end{figure}

In this section, we study the collective modes in the Landau FL
 phase as the Pomeranchuk QCP is approached, 
$0>F_l^a>-2$ in 2D and $0>F_l^a>-3$ in 3D.
These collective modes are the high partial wave channel
counterparts of the paramagnon modes in the $^3$He system.
The picture of collective modes in the $p$-wave channel at 2D
is depicted in Fig. \ref{fig:disorder}.

For this analysis, it is more convenient to employ the path integral formalism,
and perform the Hubbard-Stratonovich transformation to decouple the 4-fermion
interaction term of the Hamiltonian presented in Eq.\eqref{eq:Ham}.
After integrating out the fermionic fields, we arrive at the effective action
\bea
&&\!\!\!\!\!\!\! S_{\rm eff}(\vec n_{b})=-\frac{1}{2} \int_0^\beta d\tau \int d \vec{r}
d\vec{r}^{\; \prime} \; (f^a_l)^{-1}(\vec{r}-\vec{r}^\prime) 
\vec n_{b}(\vec{r}) \cdot \vec n_{b}(\vec{r}^\prime)
\nonumber\\
&&
+ \tr \ln \Big \{ \frac{\partial}{\partial \tau} +\epsilon(\vec{\nabla})
- \vec n_{b}(\vec{r})\cdot \vec \sigma g_{l,b} (-i\vec \nabla)
\Big \}. \nonumber \\
\label{eq:action}
\eea
In the normal FL state, we set $\bar n=0$, the 
fluctuations at the quadratic level are given by the effective action
\bea
S^{(2)}_{FL}(n)=&&\!\!\!\!\!\! \frac{1}{2 V\beta} \sum_{\vec{q}, i\omega_n}
n^{\mu a}(\vec q, i\omega_n) \; L_{\mu a, \nu b}^{(FL)} (\vec{q}, i\omega_n)\;
\nonumber \\
&\times&
n^{\nu b}(-\vec q, -i\omega_n) 
\eea
where we have introduced the fluctuation kernel $L_{\mu a, \nu b}(\vec q, i\omega_n)$ which is given by
\begin{widetext}
\bea
&&L_{\mu a, \nu b}^{(FL)} (\vec{q}, i\omega_n)=
-(f^a_1)^{-1}(q) \delta_{\mu \nu}\delta_{a b} +
\avg{Q^{\mu,a}(\vec q,i \omega_n) Q^{\nu,b}(-\vec q,-i\omega_n)}_{FL}
\nonumber \\
&&
\eea
and
\bea
&&\avg{Q^{\mu,a}(\vec q,i \omega_n) Q^{\nu,b}(-\vec q,-i\omega_n)}_{FL}=
\frac{1}{V\beta} \sum_{\vec{k}, i\omega_{n^\prime}}
\tr \{ G^{(FL)}( \vec k+\vec q, i \omega_{n^\prime}+i \omega_n)
\sigma^\mu  g_a (\hat k) G^{(FL)}( \vec k, i \omega_{n^\prime})
\sigma^\nu   g_b (\hat k)  \}. \nonumber\\
&&
\eea
\label{eq:bubble-FL}
\end{widetext}
is the correlation function of the $Q^{\mu,a}$ operators, defined in Eq.\eqref{eq:Q-momentum}, in the FL phase ({\it i.e.\/} a fermion bubble). Here
\be
G^{(FL)}(\vec k,i\omega_n)=\frac{1}{i\omega_n-\epsilon(\vec k)}
\ee
 is the fermion Green function in the FL phase, and it is diagonal in spin space.
 
After performing the Matsubara frequency summation, we find that the fluctuation kernel in the FL phase is given by the expression
\bea
&&L_{\mu a,\nu b}^{(FL)}(q, \omega)=-(f^a_l)^{-1}(q)
\delta_{\mu \nu } \delta_{ab}  
\nonumber \\
&+& 2 \delta_{\mu\nu}\int \frac{d^d k}{(2\pi)^d} \frac{n_f(\epsilon_{k-q/2})-
n_f(\epsilon_{k+q/2})}{ \omega +i\eta +\epsilon_{k-q/2}-\epsilon_{k+q/2}}
A_{ab},\nonumber \\
&&
\label{eq:collect}
\eea
where we have performed an analytic continuation to real frequency $\omega$,
and $A_{ab}$ is an angular form factor to be defined below.

\subsection{Two Dimensions}
In 2D, without loss of generality we can choose the direction of
$\vec q$ along the $x$-axis where the  azimuthal angle $\theta=0$.
The fluctuation kernels are given by
\bea
&&L_{\mu a, \nu b}^{(FL)}(q, \omega)= \delta_{ab} \delta_{\mu\nu}
\Big\{ \kappa q^2 + \delta \nonumber \\
&&+\frac{N(0)}{2} \int_{-1}^{1} \frac{d\theta}{2\pi} 
\frac{s} {s+i \eta-\cos \theta}
A^{ab}\delta_{ab} \Big \}, \nonumber \\
&&\!\!\!\!s=\omega/(v_F q), \; \delta=N(0) \left(\frac{1}{|F^a_1(0)|} -\frac{1}{2}\right)>0,
\eea
where, as before, $\mu, \nu=x,y,z$ are the components of the spin vector, and $a, b=1,2$ are the two orbital components. The diagonal components of the
angular form factors are $A^{aa}=(\cos^2 l\theta, \sin^2 l\theta)$.
For $s<1$, the angular integral can be performed to yield
\bea
&&\int^{2\pi}_0 \frac{d \theta}{2\pi}
\frac{s} {s+i \eta-\cos \theta} 
\left( \begin{array}{c}
\cos^2 l \theta\\
\sin^2 l \theta
\end{array} 
\right)\nonumber \\
&=& \frac{s}{2i \sqrt{1-s^2}}
\big\{1\pm (s-i\sqrt {1-s^2})^{2l}\big\}.
\eea
For $s\ll 1$, we can expand the above integral, and find,
for $l$ odd,
\bea
L_{\mu a, \nu b}^{(FL)}(q,\omega)&=&\delta_{ab} \delta_{\mu\nu}
 \{ \kappa q^2 +\delta +N(0)\nonumber \\
&\times& 
\Big \{
\begin{array}{lc}
-l s^2 -i l^2 s^3,& \textrm{for}\; a=1\\
l s^2 - i s, & \textrm{for}\; a=2 \\
\end{array}.
\eea
Here, since we have chosen $\vec q$ along the $x$ axis, the component $a=1$ denotes  the longitudinal component (parallel to the direction of propagation $\vec q$) and $a=2$ is the transverse component.
Similarly, for $l$ even, we get
\bea
L_{\mu a, \nu b}^{(FL)}(q,\omega)&=&\delta_{ab} \delta_{\mu\nu}
 \{ \kappa q^2 +\delta +N(0)\nonumber \\
&\times&  
\Big \{
\begin{array}{lc}
l s^2 - i s, & \textrm{for}\; a=1 \\
-l s^2 -i l^2 s^3, & \textrm{for}\; a=2\\
\end{array}.
\eea

When $l$ is odd, the transverse modes are overdamped and the longitudinal
modes are underdamped.
For example, in the case of $l=1$,
the dispersion relation at the critical point $\delta\rightarrow
0^+$ can be solved as
\be
\omega_{2}(q)=- i \frac{ \kappa v_F q^3}{N(0)}, 
\ee
for the transverse mode, and 
\be
\omega_{1}(q)=\sqrt\frac{\kappa}{N(0) } v_F q^2
-i\frac{\kappa}{2 N(0)} v_F q^3.
\ee
for the longitudinal mode.
In contrast, when $l$ is even, such as the instability in the $F^s_2$ 
channel, the transverse part is under-damped and the longitudinal part 
is overdamped \cite{oganesyan2001}.
The difference is due to the different behavior of
the angular form factors for  $l$ even and $l$ odd 
respectively.

In both cases, just as it was found in Ref. \cite{oganesyan2001}, 
the dynamic critical exponent is $z=3$.
By the power counting, the bare scaling dimension of the quartic terms
in the GL free energy, 
with effective coupling constants $v_1$ and $v_2$,
is $(d+z)-4$ where $d$ is the spatial
dimension, while the scaling dimension of the $\gamma_2$ term linear in spatial
derivatives is $(d+z)/2-2$.
All of these operators are irrelevant at zero temperature 
in 2D and 3D.
Thus, the critical theory is Gaussian, at least in perturbation theory.
However, it is possible that the above naive scaling dimensional analysis
may break down at the quantum critical point.
Various authors have found non-analytic corrections to Fermi liquid
quantities at the ferromagnetic quantum critical point 
\cite{belitz2005}.
This may also occurs here as well.
We will defer a later research for the study of these effects.
At finite temperatures, the critical region turns out
to be non-Gaussian \cite{millis1993}.
Both the terms whose couplings are $\gamma_2$ and $v_{1,2}$ now become relevant.
The relevance of the $\gamma_2$ term does not appear in the
usual ferromagnetic phase transitions \cite{lohneysen2006},
and  we will also defer the discussion
to this effect to a future publication.

\subsection{Three Dimensions}

In 3D, we can choose the $z$-axis along the direction of $\vec q$.
The diagonal part of 
the angular form factors is now $A^{aa}=(\frac{\sin^2 \theta}{2}, 
\frac{\sin^2 \theta}{2}, \cos^2 \theta)$.
Assuming that $|q| \ll k_F$,
the fluctuation kernel can be approximated as
\bea
&&L_{\mu a, \nu b}^{(FL)}(q, \omega)= \delta_{ab} \delta_{\mu\nu}
\Big\{ \kappa q^2 + \delta \nonumber \\
&&+\frac{N(0)}{2} \int_{-1}^{1} d\cos \theta 
\frac{s} {s+i \eta-\cos \theta}
A^{aa} \Big \}, \nonumber \\
&& s=\omega/(v_F q), \;
\delta= N(0) \left(\frac{1}{|F^a_1(0)|} -\frac{1}{3} \right)>0,
\nonumber \\
&&
\label{eq:3dkernel}
\eea
where $\mu, \nu=x,y,z$ are once again the three components of the spin vector. For the $l=1$ ($p$-wave) case $a, b=1,2,3$ are the three orbital components.
Using the formula
\bea
\ln \left(\frac{s+1}{s+i\eta -1}\right)=\ln \left\vert \frac{s+1}{s-1}\right\vert
-i \pi \Theta(s<1),
\eea
we arrive at
\bea
L_{\mu a, \nu b}^{(FL)}(q,\omega)&=&\delta_{ab} \delta_{\mu\nu}
 \{ \kappa q^2 +\delta\nonumber \\
&&\!\!\!\!\!\! \!\!\!\!\!\!\!\!\!\!\!\!\!\!\!\!\!\!+
\Big \{
\begin{array}{lc}
N(0)  ( s^2 -i\frac{\pi}{4}  s), & \textrm{for}\; a=1,2\\
-N(0) ( s^2 + i \frac{\pi}{2} s^3) & \textrm{for} \; a=3 \\
\end{array} ,
\eea
where only the leading order contribution to
the real and imaginary parts are kept.
The dispersion relation at the critical point $\delta\rightarrow
0^+$ can be solved as
\be
\omega_{1,2}(q)=- i \frac{4 v_F}{\pi} \frac{ \kappa q^3}{N(0)}
\label{eq:trans}
\ee
for the transverse modes, and 
\be
\label{eq:long}
\omega_{3}(q)=\sqrt\frac{\kappa}{N(0) } v_F q^2
-i\frac{\pi}{4} \frac{\kappa}{N(0)} v_F q^3.
\ee
for the longitudinal mode.
Similarly to the case in 2D, the longitudinal channel is weakly damped
and other two transverse channels are over-damped.
Again the dynamic critical exponent $z=3$, thus naively the
critical theory is Gaussian at the zero temperature.

\section{The Goldstone modes in the $\alpha$-phase}
\label{sec:alpha}

At the RPA level, the Gaussian fluctuations around the mean field
saddle point of the $\alpha$-phase are described by an effective 
action of the form
\bea
S^{(2)}_\alpha(n)= \frac{1}{2 V\beta} \sum_{\vec{q}, i\omega_n}
\delta n^{\mu a}~ L_{\mu a, \nu b}^{(\alpha)} (\vec{q}, i\omega_n)~
\delta n^{\nu b}. 
\eea
The fluctuation kernel in the $\alpha$-phase is
\bea
L_{\mu a, \nu b}^{(\alpha)} (\vec{q}, i\omega_n)&=&
-(f^a_1)^{-1}(q) \delta_{\mu \nu}\delta_{a b} +
\avg{Q^{\mu,a}(\vec q, i \omega_n) \nn \\
&\times&
Q^{\nu, b}(-\vec q,-i\omega_n)}_\alpha,
\eea
where
\begin{widetext}
\be
 \avg{Q^{\mu,a}(\vec q, i \omega_n) Q^{\nu, b}(-\vec q,-i\omega_n)}_\alpha=
\frac{1}{V\beta} 
\sum_{\vec{k}, i\omega_{n^\prime}} \tr \{ G^{(\alpha)}(\bar{n}, k+q, i \omega_{n^\prime}+i \omega_n)
\sigma^\mu g_{l,a}(\hat k) 
G^{(\alpha)}(\bar{n}, k, i \omega_{n^\prime}) 
\sigma^\nu   g_{l,b} (\hat k)  \}
\ee
\end{widetext}
is the correlator of the operators $Q^{\mu a}$ in the mean field 
theory ground state of the $\alpha$-phase (again a fermion bubble).
Here $G^{(\alpha)}(\bar{n}, \vec k, i \omega_{n})$ is the fermion 
propagator in the $\alpha$-phase with an expectation value of the 
(nematic-spin-nematic for the $l=2$ case) order parameter equal to $\bar n$,
\be
G^{(\alpha)}(\bar{n}, \vec k, i \omega_{n}) =\left(i\omega_n-\epsilon_\alpha(\vec k,\bar n)\right)^{-1}
\ee
where
\bea
&&\epsilon_\alpha(\vec k,\bar n)=\epsilon(\vec k)-\avg{\vec n_b}_\alpha \cdot \vec \sigma \; g_{l,b}(\hat k) \nonumber \\
&& g_{l,1}(\vec k)=\cos l \theta_{\vec k} , 
\quad g_{l,2}(\vec k)=\sin l \theta_{\vec k},
\eea
is the fermion dispersion, a matrix in spin space, and $\avg{\vec n_b}_\alpha =\bar n \; \hat z \; \delta_{b,1}$ is the mean field expectation value of the order parameter in the $\alpha$-phase.

Since in the $\alpha$-phase there are spontaneously broken continuous 
symmetries, both in 2D and in 3D, the collective modes will consist 
of gapped longitudinal modes, {\it i.e.\/} along the direction of the 
condensate, and gapless, Goldstone, modes transverse to the
direction of spontaneous symmetry breaking,
both in the density and spin channels.
We will discuss the Goldstone modes in the $\beta$-phase in the next
section.

We next comment on the stability of the $\alpha$-phase in the 
$p$-wave channel.
The GL energies Eq.\eqref{eq:GL2} and Eq.\eqref{eq:3Dinstab} contain a cubic
term linear in derivatives.
In the ordered state, it might induce a linear derivative coupling 
between the massless Goldstone mode at the quadratic level through the
condensate longitudinal mode, thus leading to a Lifshitz instability
in the ground state. As we will show, this indeed occurs in the
$p$-wave $\beta$-phase.
However, we will see that in the $\alpha$-phase
the spin (density) channel Goldstone modes have
the same orbital (spin) indices as those of the longitudinal mode,
thus they can not be coupled by Eq. \eqref{eq:GL2} and Eq.\eqref{eq:3Dinstab}.
Instead, the Goldstone modes couple to other gapped modes
at the quadratic level through linear derivative terms, which 
do not lead to instability at weak coupling, but can
renormalize the stiffness of the Goldstone modes.

\subsection{The 2D $\alpha$-phases}
\label{subsec:2dalpha}
\begin{figure}
\psfrag{Q}{$\vec K$}
\psfrag{q}{$\vec q$}
\psfrag{x}{$x$}
\psfrag{y}{$y$}
\psfrag{phi}{$\phi$}
\begin{center}
\includegraphics[width=0.35\textwidth]{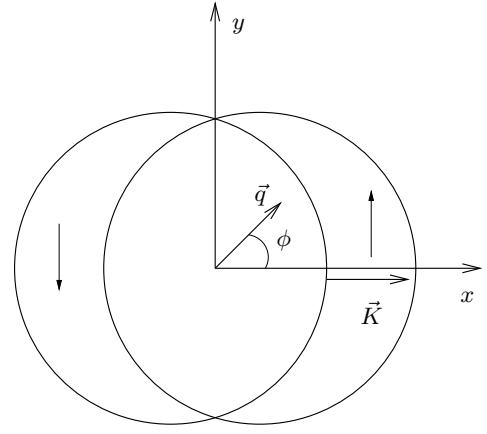}
\end{center}
\caption{The azimuthal angle $\phi$ for the propagation wavevector
$\vec q$ for the Goldstone modes in the 2D alpha phase.
The spin down and up Fermi surfaces can overlap each other by
a shift at the wavevector $\vec K= 2\bar n/v_F$.
}\label{fig:GSalpha2D}
\end{figure}

We consider the 2D $\alpha$-phases assuming 
the order parameter configuration as $\avg{n_{\mu b}}=\bar n
\delta_{\mu z}\delta_{b 1}$.
The order parameter is thus the operator $n_{z,1}$ of
Eq. \eqref{eq:order-parameters} and Eq. \eqref{eq:g12}.
As studied in Sec. \ref{subsec:alpha}, the Goldstone mode manifold
$S^2\times SO_L(2)$ results in one branch of Goldstone mode
in the density channel, and two branches of Goldstone modes in the spin channel.The fluctuation kernel of the $\alpha$-phases, $L^{(\alpha)}_{\mu a,\nu b}$ is a $6 \times 6$ matrix. Its eigenvalues will thus yield 
six collective modes of which three are the above mentioned Goldstone modes. 
The other three modes are gapped, and are associated with 
the structure of the order parameter in the $\alpha$-phases. 
In the low frequency regime $\omega\ll \bar n$, we can neglect
the mixing between the Goldstone modes and other gapped modes.

The density channel Goldstone mode is associated with the field $n_{z,2}$, conjugate to the bilinear fermion operator $Q_{z,2}$, is longitudinal in the spin sector and transverse in the charge sector,
\bea
Q_{z2}(\vec r)= \psi^\dagger_\alpha(\vec r) 
[\sigma_{z,\alpha\beta} (-i\nabla_y)] \psi_\beta(\vec r),
\eea
and describes the Fermi surface oscillation in the
$2$-direction while keeping the spin configuration unchanged. On the other hand, the spin channel Goldstone modes $n_{sp,x\pm iy}$, conjugate to 
the fermion bilinears $Q_{x \pm i y,1}=(Q_{x,1} \pm i Q_{y,1})/2 $, 
describe spin oscillations 
while keeping the Fermi surface unchanged. 

\subsubsection{Density channel Goldstone mode}

The density channel Goldstone field $n_{z,2}$
behaves similarly to its counterpart in the density channel
Pomeranchuk instability \cite{oganesyan2001}.
The same approximation  as in Ref. \cite{oganesyan2001}
can be used to deal with the anisotropic Fermi surface, {\it i.e.\/},
keeping the anisotropy effect in the static part of the correlation function, 
but ignore it in the dynamic part.
This approximation is valid at small values of order parameter,
{\it i.e.\/}, $x=\bar n /(v_F k_F)\ll 1$, where $\bar n= \langle n_{z,1} \rangle$.
We define the propagation wavevector $\vec q$ with the azimuthal angle
$\phi$ as depicted in Fig. \ref{fig:GSalpha2D} for the $l=1$ case.

The effective fluctuation kernel for the charge channel Goldstone mode, 
which we label by $fs$, reduces to
\bea
L^{(\alpha)}_{fs}(\vec q,\omega)&=&\kappa q^2- N(0)
\begin{cases}
i s  \sin^2( l \phi) - ls^2  \cos(2 l\phi) \\
\qquad (\mbox{$l$ even}), \\ 
i s \cos^2( l \phi) + ls^2  \cos( 2 l\phi) \\
\qquad (\mbox{$l$ odd} ).
\end{cases}
\nn \\&&
\eea
where $s=\omega/v_F q$.
Similarly to the results of in Ref. \cite{oganesyan2001}, 
this Goldstone mode
corresponding to Fermi surface oscillation is 
overdamped almost on the entire Fermi surface except on a set of directions of measure zero:
for $l$ even, the charge channel Goldstone mode is underdamped in the directions
$\phi=n\pi/l$, $n=0,1,\ldots,l$, which are just the symmetry axes of the Fermi surface; for $l$  odd, it is underdamped instead in the directions $\phi=\pi(n+1/2)/l$, and in this case the Goldstone mode
is maximally damped along the symmetry axes.

\subsubsection{Spin channel Goldstone modes}

On the other hand, the spin channel Goldstone fields $n_{x\pm i y,1}$
behave very differently. They only involve ``inter-band transitions'', leading instead to a fluctuation kernel of the form
\bea
&&L_{x+iy,1}(\vec q,\omega)= \kappa  q^2 +\frac{1}{|f^a_1|}
+2 \int \frac{d^2 k}{(2\pi)^2} \cos^2 (l \theta_k) \nn  \\
&\times&
\frac{ n_f [\xi_\downarrow(\vec k-\vec q/2)]-n_f [\xi_\uparrow(\vec k+\vec q/2)]}{\omega+i\eta
+\xi_\downarrow(\vec k-\vec q/2)-\xi_\uparrow(\vec k+\vec q/2)}, \ \ \
\label{eq:gsspinalpha}
\eea
where 
$\xi_{1,2}(k)=\epsilon(\vec k)-\mu\mp \bar n \cos (l \theta_k)$.
They satisfy the relation of
\bea
L_{x-iy,1}(\vec q, \omega)=
L_{x+iy,1}(-\vec q, -\omega).
\eea
A detailed calculation, presented in Appendix
\ref{app:gsspinealpha}, shows that 
for $\frac{\omega}{\bar n},\frac{v_F q}{\bar n}\ll 1$,
the kernel $L_{x \pm i y,1}$
reads
\bea
L_{x \pm iy, 1}(\vec q, \omega)=
\kappa q^2- \frac{N_0}{2|F^a_l|} \frac{\omega^2}{\bar n^2},
\eea
which gives rise to a linear and undamped spectrum:
\bea
\omega_{x \pm i y,1}(\vec q)=\sqrt{\frac{2\kappa|F^a_l|}{N(0)}} \; \bar n\; \vert \vec  q\vert.
\label{eq:x+iy-dispersion}
\eea

Eq. \eqref{eq:x+iy-dispersion} has to  two important features 
(due to the interband transition): the isotropy of dispersion relation at 
$\frac{\omega}{\bar n},\frac{v_F q}{\bar n}\ll 1$
in spite of the anisotropic Fermi surfaces, and the underdamping
of the Goldstone modes.
The contribution to the integral comes from the region around 
Fermi surfaces with the width about $2\bar n\cos( l\theta_k)/v_F$.
The dependence of the integral on $\vec q$ can be
neglected at $\frac{v_F q}{\bar n}\ll 1$ because a small $\vec q$ 
changes the integration area weakly, and it
thus matters only for high order corrections in $\vec q$.
The contribution to damping comes from the region where
two bands become nearly degenerate, {\it i.e.\/}, $\cos ( l \theta_k) \approx 0$.
However, the angular form factor also takes the
form of $\cos (l \theta_k)$, which tends to suppress damping.
As $v_F q$ becomes comparable to $\bar n$, the anisotropy and
damping effects  become more important.

The linear dispersion relation for the spin-channel Goldstone modes 
at $\frac{v_F q}{\bar n}\ll 1$ holds 
regardless of whether $l$ is  odd or even.
This fact is closely related to time reversal (TR) and parity 
symmetry properties of the order parameter $n^{\mu a}$.
For $l$ odd,  $n^{\mu a}$ is {\em even} under TR 
transformation, and hence terms linear in time derivatives
cannot not appear in the effective action.
On the other hand, for $l$ even even, although $n^{\mu a}$ is {\em odd} under TR,
in 2D we can still define the combined transformation $T^\prime$  as
\bea
T^\prime=T R(\pi/l),
\eea
under which $n^{\mu a}$ is {\em even}. 
Here $R(\pi/l)$ is a real space rotation by an angle of $\pi/l$.
thus, also in this case, terms which are linear in time derivative are not allowed
in the effective action.
In contrast, for the case of a ferromagnet at $l=0$,
TR symmetry is broken, and no other symmetry exists 
to form a combined operation $T^\prime$
that will leave the system invariant.
As a result, terms linear in time derivatives appears in the effective action of a ferromagnet. The same arguments apply for phases with mixed ferromagnetic and spin nematic order (and its generalizations).
Furthermore, in the presence of time-reversal-violating terms, the two transverse components of spin fluctuation become
conjugate to each other as in the presence of ferromagnetic long range order.
In this case, only one branch of spin wave Goldstone mode exists with a
quadratic dispersion relation $\omega_{FM}\propto q^2$.

\subsubsection{Spin wave spectra}

We assume that the $F_0^a$ channel is off-critical, thus in the
normal state no well-defined spin wave modes exist.
However, in the $\alpha$-phase the spin channel Goldstone modes carry
spin, thus  induce a well-defined pole in the spin wave
spectrum.
This can be understood from the commutation relation between 
spin modes and the spin channel Goldstone modes 
\bea
[S_x\pm iS_y, Q_{x \mp iy,1}]&=& \pm2 Q_{z,1}.
\eea
In the $\alpha$-phase where $Q_{z1}$ obtains a non-vanish expectation
value, then these two channels become conjugate.
As a result, the spin-wave gains a sharp resonance and should
exhibit in the neutron scattering experiment.
In contrast, in the normal state, the coupling between this
two modes is negligible, and thus the resonance disappears.
A similar physics occurs in the $SO(5)$ theory for the 
explanation of $\pi$-resonance in the underdoped high T$_c$ cuprates
\cite{demler2004}.

The effective coupling constant which mixes the $S_x+i S_y$ and 
$Q_{x-iy,1}$ operators is a bubble diagram which can be
calculated as
\bea
\chi^0_s(\vec q,\omega)=-N(0) \frac{\omega}{\bar{n}}.
\eea
This bubble is dressed by the interaction in the $F^a_l$ channel.
The resonant part of the spin correlation function ({\it i.e.\/} 
the contribution of the collective mode pole) becomes
\bea
\chi_s(\vec q,\omega)&=&\avg{S_+(\vec q, \omega) S_- (-\vec q,-\omega)}=
\frac{|\chi^0_s (\vec q,\omega)|^2}{L_{x+iy,1}(\vec q,\omega)}\nn \\
&=&\frac{ N(0)  \frac{\omega^2}{\bar{n}^2}  }
{\frac{\kappa q^2}{N(0)}  -\frac{2}{|F^a_l|} 
\frac{\omega^2}{\bar{n}^2}  -i \delta}.
\eea
For fixed but small $\vec q$, 
the spectral function exhibits a $\delta$-function peak at the
dispersion of the collective mode
\bea
\textrm{Im} \chi_s (\vec q,\omega)= \kappa \pi
v^2_F q^2 \bar n^2 |F^a_1|^2 \delta (\omega^2-\omega_q^2).
\eea
which will induce a spin resonance in all
directions. It is worth to note that in the spin channel the isotropy 
in this dispersion relation at small $\vec q$ persists even deeper 
in the ordered phase.

\subsection{ The 3D $\alpha$-phase for  $l=1$ }

In the 3D $\alpha$-phase, we assume the Fermi surface distortion
along the $z$-axis, and the order parameter configuration as
$n^{\mu a}=\bar n \delta_{\mu, z} \delta_{a,3}$.
Similarly to the 2D case,  the spin up and down Fermi surfaces
are related by a overall shift at the wavevector $\vec K=2\bar n/v_F \hat z$.
The remaining symmetry is $SO_L(2)\otimes SO_S(2)$
which results in four Goldstone modes.
They can be classified as two density channel
modes and two spin channel modes.
Without loss of generality, we choose the propagation wavevector
$\vec q$ lies in the $xz$-plane.

The density channel Goldstone modes describe the Fermi surface 
oscillations
in the $x$ and $y$ directions, which are associated with
the fields $n_{z,1}$ and $n_{z,2}$.
By a Legendre transformation,
they are conjugate to the bilinear operators $Q_{z,1}$
and $Q_{z,2}$:
\bea
Q_{z1}(\vec r)&=& \psi^\dagger_\alpha(\vec r) 
[\sigma_{z,\alpha\beta} (-i\nabla_x)] 
\psi_\beta(\vec r)
\nonumber \\
Q_{z2}(\vec r)&=&\psi^\dagger_\alpha(\vec r) 
[\sigma_{z,\alpha\beta} (-i\nabla_y)] \psi_\beta(\vec r)
\eea
Following the same procedure of the calculation in 2D, 
we find that the fluctuation kernel of the Goldstone mode 
$n_{z,1}$ is
\bea
L_{z1}=\kappa q^2- i \frac{\pi}{4} N(0)  s \cos^2 \theta_q
+N(0) s^2 \cos 2 \theta_q .
\eea
It is overdamped almost everywhere, except if $\vec q$ lies in 
the equator ($\theta=\pi/2$), in which case it is underdamped and has a quadratic dispersion.
On the other hand, the fluctuation kernel of the
Goldstone mode of $n_{z,2}$ reads
\bea
L_{z,2}&=&\kappa q^2 - i N(0) s \frac{\pi}{4}, 
\eea
which has no dependence on the angle $\theta_q$. Hence, this mode
is over-damped on the entire Fermi surface.

The spin channel Goldstone modes of $n_{x\pm iy, 3}$
in the $F^a_1$ channel behaves similarly to 
that in the 2D case.
We simply present their fluctuation kernels at small 
wavevectors as
\bea
L_{x\pm i y,3}(\vec q, \omega)
&=&\kappa q^2 -\frac{3 N(0)}{|F_1^a|} s^2\
\label{eq:gsspinalpha3}
\eea
where  $s=\omega/ v_F  |K|$.
The spin wave excitation is also dressed by the interaction in
the $F^a_1$ channel in the $\alpha$-phase.
By a similar calculation to the 2D case, we have
\bea
\chi_s(\vec q,\omega)&=&\avg{S_+(\vec q, \omega) S_- (-\vec q,-\omega)}\nonumber \\
&=&
\frac{|\chi^0_s (\vec q,\omega)|^2}{L_{x+iy,3}(\vec q,\omega)}.
\eea
Thus it also develops the same pole as
in the spin channel Goldstone modes.

\section{Goldstone modes in the  $\beta$-phases }
\label{sec:beta}
In this section, we calculate the Goldstone modes and spin wave spectra
in the $\beta$-phase at 2D and 3D at the RPA level.
We will show that for $l=1$, a Lifshitz-like instability arises
leading to a spatially  inhomogeneous ground state.
This is because of a dynamically generated Dzyaloshinskii-Moriya 
interaction among the Goldstone modes as a result of the spontaneously
breaking of parity.
A G-L analysis is presented to analyze this behavior.

\subsection {The 2D $\beta$-phases}
\begin{figure}
\psfrag{x}{$x^\prime$}
\psfrag{y}{$y^\prime$}
\psfrag{z}{$z$}
\psfrag{phi}{$\phi$}
\psfrag{q}{$\vec q$}
\psfrag{s}{$\vec s$}
\psfrag{l}{$l=1,w=1$}
\begin{center}
\includegraphics[width=0.35\textwidth]{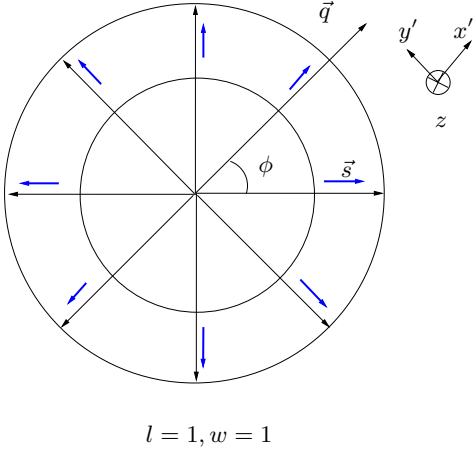}
\end{center}
\caption{The three Goldstone modes of
$O_{z}, O_{x^\prime}, O_{y^\prime}$ 
of the 2D $\beta$-phase can be viewed as a relative
spin-orbit rotation for this phase with angular momentum $l=1$ and winding number $w=1$. Here
$\phi$ is the  azimuthal angle of the propagation direction $\vec q$.
}\label{fig:goldstone1}
\end{figure}   

\begin{figure}
\psfrag{o}{$\omega=v_{gs} q$}
\psfrag{p}{$q^\prime$}
\psfrag{1}{$2\bar n$}
\psfrag{2}{$2\bar n +v_F q$}
\psfrag{3}{$2\bar n -v_F q$}
\psfrag{q}{$q$}
\psfrag{e}{$\omega$}
\psfrag{particle-hole}{\small particle-hole}
\psfrag{continuum}{\small continuum}
\begin{center}
\includegraphics[width=0.35\textwidth]{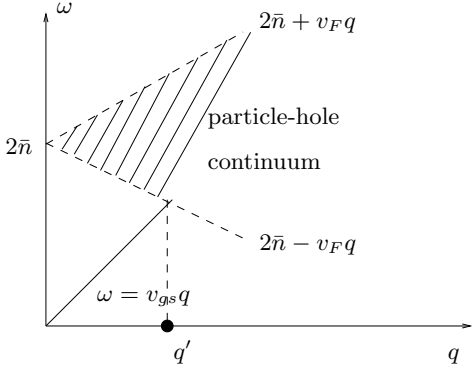}
\end{center}
\caption{The linear dispersion relation of Goldstone modes  of the 2D $\beta$-phases ($l\ge 2$) at 
small momentum $\vec q$. For $ q> q^\prime$, the Goldstone
modes enter the particle-hole continuum, and are damped.
}\label{fig:GSdispersion1}
\end{figure} 

Without loss of generality, we consider the $\beta$-phase in the $F^a_l$
channel.
We first assume a uniform ground state with the configuration of 
the $d$-vector $\vec d =(\cos l\theta,\sin l\theta,0)$ as defined 
in Eq. \eqref{eq:dvector}.
The corresponding order parameter {\it i.e.\/}, the Higgs mode
$n_{\rm higgs}$, is conjugate to operator $O_{higgs}$ as
\bea
O_{\rm higgs}(\vec r)=\frac{1}{\sqrt 2}\big( Q_{x,1}(\vec r)+Q_{y,2}(\vec r)
\big).
\eea
By performing a relative spin-orbit rotation around the $z$, $x$, 
and $y$-axes  on the mean field  ansatz, we obtain the operators 
for three branches of Goldstone modes as
\bea
O_z(\vec r)&=&\frac{1}{\sqrt 2} ( Q_{x,2}(\vec r)-Q_{y,1}(\vec r)),
 \nonumber \\
O_x(\vec r)&=& - Q_{z,2}(\vec r), \ \ \  O_y(\vec r)= Q_{z,1}(\vec r).
\label{eq:goldstone}
\eea

We define the propagation wavevector of Goldstone modes $\vec q$
and its azimuthal angle $\phi$ as depicted in Fig. \ref{fig:goldstone1}.
For the general direction of $\vec q$, it is more convenient to
set up a frame with three axes $x^\prime$, $y^\prime$ and $z$,
where $x^\prime\parallel \vec q$ and $y^\prime\perp \vec q$.
We rotate $O_x$ and $O_y$ into
\bea
O_{x^\prime}&=& \cos l\phi ~O_{x} -\sin l\phi~ O_{y}
\nonumber \\
O_{y^\prime}&=& \sin l\phi~ O_{x} +\cos l\phi ~O_{y}.
\eea
$O_{z}$, $O_{x^\prime}$, and $O_{y^\prime}$ are the generators of a relative
spin-orbit rotation around the $z$, $x^\prime$, and $y^\prime$-axis
respectively.
Thus, in the following, we call the Goldstone mode of
$O_{x^\prime}$ the longitudinal Goldstone
mode, and those of $O_z$ and $O_{y^\prime}$ are
two transverse Goldstone modes.

The system with $\vec d=(\cos l\phi,\sin l\phi,0)$ has the
following reflection symmetry even in the presence of the
$\vec q$,
\bea
&&\theta_k\rightarrow 2\phi -\theta_k, \ \ \
\sigma_z\rightarrow -\sigma_z, \nonumber \\
&&\sigma_x\rightarrow \sigma_x \cos 2l \phi +\sigma_y \sin 2l \phi,
 \nonumber \\
&&\sigma_y\rightarrow \sigma_x \sin 2l \phi -\sigma_y \cos 2l \phi.
\eea
$O_{z}$ and $O_{y^\prime}$ are even under this transformation
while $O_{x^\prime}$ is odd.
Thus, $O_{x^\prime}$ decouples from $O_{y^\prime}$ and $O_z$,
while hybridization occurs between $O_{z}$ and $O_{y^\prime}$.
For a small wavevector $\frac{v_F q}{\bar n} \ll 1$
and low frequency $\frac{\omega}{\bar n}\ll 1$,
we can ignore the mixing between the Goldstone modes
and other gapped modes.
For $l \ge 2$, the eigenvalues of the fluctuation kernel for the Goldstone modes is
\bea
L_{z z} (q,\omega)&\approx& L_{x^\prime x^\prime} (q,\omega)
\approx  L_{y^\prime y^\prime} (q,\omega)\nonumber \\
&\approx& \kappa q^2 -\frac{\omega^2}{4 \bar n^2} \frac{N(0)}{|F^a_l|},
\eea
where we have neglected the anisotropy among the three dispersion relations.
A finite hybridization between $O_z$ and $O_{y^\prime}$ 
appears at the order of $O(q^l)$
\bea
L_{zy^\prime} (q,\omega) \propto i q^l.
\label{eq:RPA}
\eea
which is negligible at small $q$ at $l>2$.
Thus, the spectrum of the Goldstone modes is linear for $l>2$.
For $l=2$, the hybridization is quadratic in $\vec q$
\bea
L_{zy^\prime} (q,\omega)= -i \frac{N(0)}{32\sqrt{2} k_F^2} 
q^2 (1+4b) \ll \kappa q^2,
\eea 
and thus must be taken into account.
The resulting eigenmodes in the transverse channel 
are $O_z\pm iO_{y^\prime}$.
However, the linear dispersion relation remains at $l=2$.

Similarly to ferromagnets, in there are two Fermi surfaces with 
unequal volume in the $\beta$-phases.
The interband transition has a gap of $2\bar n$ and a particle-hole
continuum of width $2v_F q$ as depicted in Fig. 
\ref{fig:GSdispersion1}.
The Goldstone modes correspond to the interband transition
with a velocity  $v_{gs}\simeq 2\bar n \sqrt{\kappa |F^a_l(0)|/N(0)}$,
and no Landau-damping effects exist at small $q$.
Naturally, after Goldstone modes enters
the  particle-hole continuum, at the wavevector 
$q^\prime\approx 2\bar n/(v_{gs}+v_F)$,
the mode is no longer long lived and become Landau damped.

The linear dispersion relation of the Goldstone modes  holds for all the values
at $l\ge 2$.
This feature is also due to the symmetry properties of the $n^{\mu a}$
under TR and parity transformation.
The reasoning here is the same as that for the $\alpha$-phase
in the Section \ref{subsec:2dalpha}.

We next calculate the spin-wave spectra in the $\beta$-phase
at $l\ge 2$. We have the following commutation relations as
\bea
&&[S_+ e^{i\phi_q}, e^{-i\phi_q} (O_x-i O_y)/\sqrt 2]=
i O_{hig} + O_{z}, \nn \\
&&[S_z, O_z]= i O_{hig}.
\eea
The effective coupling between $S_z$ and $O_z$ can be calculated as
\bea
\chi^0_{Sz}(\vec q,\omega)
=\frac{-i}{\sqrt 2} \frac{\omega}{\bar n} \frac{N(0)}{|F^a_l|}
+O(q^2).
\eea
Thus the spin-spin correlation function, dressed by the $F^a_l$
channel interactions,  near the resonance of the dispersing transverse collective mode has the form
\bea
\chi_{s}(\vec q,\omega)&=&\avg{S_z(\vec q,\omega) S_z(-\vec q,-\omega)}
=\frac{|\chi^0_{s}|^2}{L_{zz}(\vec q, \omega)}\nn \\
&=& \frac{2 N(0)}{|F^a_l|} \frac{\omega^2}{\omega^2-4 \bar n^2
\frac{|F^a_l|}{N(0)} \kappa q^2}.
\nonumber \\
&&
\eea
The spectral function at the resonance reads
\bea
\textrm{Im} \chi_s(\vec q,\omega)= 8 \bar n^2 \kappa q^2 \delta(\omega^2-\omega_q^2).
\eea
Similarly, the effective coupling between $S_+ e^{i\phi_q}$ and  
$e^{-i\phi_q} (O_x-i O_y)/\sqrt 2$ gives the same result,
\be
\chi^0_{s_{x\pm iy}}=\chi^0_{sz}
\ee
The transverse spin-spin correlation function is
 \be
 \avg{S_+(\vec q, \omega)
S_-(-\vec q, -\omega)}= \frac{|\chi^0_{s_{x\pm iy}}|^2}{ L_{x+iy, x-iy}}
\ee 
which is also dressed by the interactions.

\subsection{Lifshitz-like instability in the 2D $p$-wave channel}
\label{subsec:lifshitz}

\begin{figure}
\psfrag{Q}{$\omega^2$}
\psfrag{O}{$O_{z-iy^\prime}$}
\psfrag{S}{$O_{x^\prime}$}
\psfrag{T}{$O_{z+i y^\prime}$}
\psfrag{q}{$q$}
\psfrag{q'} {$q_c$}
\psfrag{2nh}{$2\bar n |f^a_1| h_{so}$} 
\begin{center}
\includegraphics[width=0.3\textwidth]{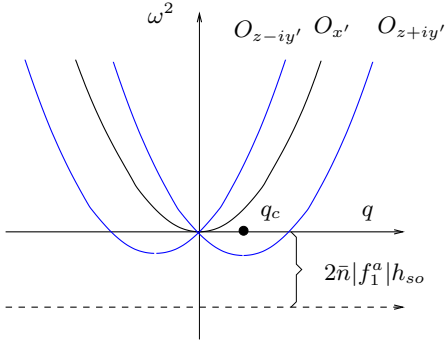}
\end{center}
\caption{[color online] The dispersion relation of Goldstone modes for the $\beta$-phase with $l=1$.
The longitudinal modes $O_{y^\prime}$ has linear dispersion relation,
while the two transverse modes $O_{z\pm i y^\prime}$ are unstable
towards the Lifshitz-like instability at small momentum $\vec q$.
}\label{fig:GSdispersion2}
\end{figure}

\begin{figure}
\psfrag{x}{$\hat e_x$}
\psfrag{y}{$\hat e_y$}
\psfrag{z}{$\hat e_z$}
\psfrag{1}{$\vec n_1$}
\psfrag{2}{$\vec n_2$}
\psfrag{3}{$\vec n_3$}
\psfrag{A}{\small A}
\psfrag{B}{\small B}
\begin{center}
\includegraphics[width=0.4\textwidth]{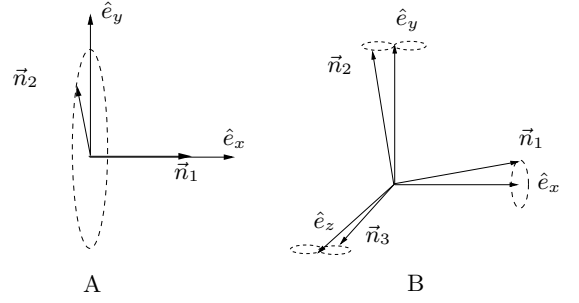}
\end{center}
\caption{Longitudinal and transverse twists for the Lifshitz-like
instability in the $\beta$-phase with $l=1$.
A) The longitudinal twist with $\hat n_2$  processing around the
$\hat e_1$ axis as described in Eq. \eqref{eq:ltwist}.
B) The transverse twist with the triad formed by $\hat n_{1}$, $\hat n_{,2}$, and $\hat n_3=
\hat n_1 \times \hat n_2$ precessing around the $\hat e_x$ axis
as described in Eq. \eqref{eq:trtwist}.
}\label{fig:lifshitz}
\end{figure}

For $l=1$, with the assumption of the uniform ground state,
the dispersion for the longitudinal mode $O_z$ remains linear as
\bea
\omega_{L}^2 =4 \bar n^2 |F_1^a| \frac{\kappa}{N(0)} q^2.
\eea
However, the situation for the transverse modes is dramatically different.
The mixing between $O_z$ and $O_{y^\prime}$ scales linearly with $q$:
\bea
L_{zy^\prime} (\vec q,\omega)=- L_{y^\prime z} (\vec q,\omega)
\approx \frac{i}{4\sqrt 2} \left(b+\frac{3}{2}\right) N(0) 
x \frac{q}{k_F}. \nn \\
\eea
We diagonalize the matrix and then obtain the eigenmodes as
$O_{z}\pm i O_{y^\prime}$ with the following dispersion relation
\bea
\omega_{T,\pm}^2= 4 \bar n^2 |F_1^a(0)| \left(
\frac{\kappa q^2}{N(0)} \pm c  x \frac{q}{k_F}\right),
\label{eq:lifshitz}
\eea
as depicted in Fig. \ref{fig:GSdispersion2}
where $c$ is a constant at the order of 1.

Clearly, $\omega_{T,-}^2<0$ for small $|\vec q|$. This means that the uniform ground state can not be stable in the $F_1^a$ 
channel due to a  Lifshitz-like instability \cite{FETTER2004,Kim2004}.
This instability can be understood in terms of the nontrivial effects of the gradient term 
with coefficient $\gamma_2$ in the G-L free energy in Eq.\eqref{eq:GL2}. this term
is cubic in the order parameter $n^{\mu,b}$ and
a linear in spatial derivatives.
This term leads to an inhomogeneous ground state in the $\beta$-phase in which
parity is spontaneously broken.
A similar phenomenon occurs in the bent-core liquid crystal system
where a spontaneously chiral inhomogeneous nematic state
arises in an non-chiral system.
A similar term involving a linear derivative and the cubic order of
order parameter is constructed in Ref. \cite{lubensky2002}
to account for this transition.
These inhomogeneous ground states also  occur in chiral liquid
crystal \cite{chaikin1995} and helimagnets \cite{binz2006} such
as MnSi in which parity is explicitly broken.
In contrast, such a term is prohibited in the ferromagnetic transition
\cite{hertz1976,millis1993}, and the nematic-isotropic phase 
transition \cite{oganesyan2001} in the density channel Pomeranchuk instabilities.
We will see that the G-L analysis below
based on the symmetry argument agrees with the
RPA calculations in Eq.\eqref{eq:RPA}.

In order to obtain the values of $\gamma_{1,2}$, we linearize the gradient
terms in Eq. \eqref{eq:GL2} in the ground state with $d$-vector configuration
of $\vec d=(\cos \theta, \sin \theta,0)$, i.e.,
\bea
\vec n_1 &=& \sqrt{\bar n^2-\delta n_1^2} \hat e_x +\delta \vec n_1, \nn \\
\vec n_2 &=& \sqrt{\bar n^2- \delta n_2^2} \hat e_y +\delta \vec n_2,
\eea 
where $\delta \vec n_1 \perp \hat e_x$, $\delta \vec  n_2 \perp \hat e_y$
and $\vec n_1 \perp \vec n_2$ is kept.
The contribution from the Goldstone modes can be organized into
\bea
F_{grad}(\vec O)&=& \gamma_1 \{(\partial_i O_x)^2+ (\partial_i O_y)^2+
(\partial_i O_z)^2\} \nn \\
&+&\frac{\gamma_2 \bar n}{\sqrt 2} (O_x \partial_y O_z-
O_y \partial_x O_z) \nn \\
&+& \gamma_2 \bar n^2 (\partial_x O_x +\partial_y O_y).
\label{eq:DM2D}
\eea
With the assumption of the uniform ground state, the first $\gamma_2$  term
behaves like a Dzyaloshinskii-Moriya term, and gives a liner dependence
in the dispersion relation as appears in Eq.\eqref{eq:lifshitz}.
By matching the coefficients, we arrive at the result
\bea
\gamma_1=\kappa, \ \ \ \gamma_2=\frac{\sqrt{2}c N(0)}{v_F k_F^2}.
\label{eq:gradcoeff}
\eea

The second coefficient $\gamma_2$ in the GL expansion of Eq.\eqref{eq:DM2D} is a total derivative 
of the longitudinal Goldstone modes.
It does not contribute to the equation of motion for the Goldstone modes
around the saddle point of the uniform mean field ansatz.
However, Eq.\eqref{eq:GL2} actually allows a twisted ground
ground state in the longitudinal channel as depicted in
Fig. \ref{fig:lifshitz} A.
Without loss of generality, we
assume a pitch vector along the $x$ axis, $\vec q \parallel \hat x$, and
perform a longitudinal twist of the Goldstone configuration
of $\vec n_{1,2}$ as
\bea
\vec n_1 = \bar n (1,0,0),  \ \ \
\vec n_2 = \bar n (0,\cos q x, -\sin q x).
\label{eq:ltwist}
\eea
This configuration means that we fix $\vec n_1$, and
rotate $\vec n_2$ around the $x$-axis.
If the system has a small external spin-orbit coupling, it
pins the order parameter configuration.
We introduce an effective spin-orbit field $h_{so}$  to describe this effect
\bea
V(h_{so})=-h_{so} (n_{x,1}+n_{y,2}).
\eea
Then the G-L free energy becomes
\bea
F(n)=\gamma_1 \bar n^2 q^2 -\gamma_2 \bar n^3 q-h_{so} \bar n \cos qx.
\label{eq:engltwist}
\eea
If $h_{so}$ is less than a critical value $h_{so,L}$ defined as
\bea
h_{so,L}=\frac{\gamma_2^2 \bar n^3}{4\gamma_1},
\label{eq:hsol}
\eea
then the Lifshitz instability occurs with the pitch of
$q_c=\gamma_2 \bar n /( 2\gamma_1)$.
If $h_{so}>h_{so,L}$, the instability due to the longitudinal
Goldstone mode is suppressed.

Now let us look at the instability caused by the transverse Goldstone modes
as depicted in Fig. \ref{fig:lifshitz} B.
The transverse mode $O_z+iO_y$ describes the precession
of the triad of $\hat n_{1}$, $\hat n_{2}$, $\hat n_3=\hat n_1 \times \hat n_2$ as
\bea
\vec n_1&=& \bar n \big(\cos \epsilon ~\hat e_x+  \sin \epsilon \cos q x
~\hat e_y-\sin \epsilon \sin qx ~\hat e_z\big), \nn \\
\vec n_2&=& \bar n \big \{-\sin \epsilon~ \cos qx ~\hat e_x+
(\cos^2 \frac{\epsilon}{2} -\sin^2 \frac{\epsilon}{2} ~\cos 2 qx)~\hat e_y 
\nn \\
&-& \sin^2 \frac{\epsilon}{2} \sin 2 q x ~\hat e_z \big \} ,
\label{eq:trtwist}
\eea
where $\epsilon$ describes the precession amplitude;
$\vec n_1$ precesses around the $x$-axis, and
$\vec n_2$ traces a  ``figure of eight'' orbit on the surface
of the unit sphere.
To linear order of $\epsilon$,
the free energy cost is
\bea
V(n)= \frac{3}{2} \gamma_1 q^2 (\bar n \epsilon)^2 +
\frac{\bar n}{2} q \gamma_2 (\bar n \epsilon)^2
+\frac{h_{so}}{\bar n} \frac{3}{4}(\bar n \epsilon)^2.
\eea
Hence, when $h_{so}$ is larger than a critical value $h_{so,T}
=\gamma_2^2 \bar n^3/18 \gamma_1<h_{so,L}$, the instability 
due to the transverse Goldstone modes is suppressed.

Because $h_{so,L}>h_{so,T}$, the longitudinal Goldstone channel
instability is stronger than that in the transverse channel.
Thus, in the absence of the external field, the ground state
exhibits the longitudinal twist of Eq.\eqref{eq:ltwist}.
However, this spiral order can not give static Bragg peaks
in neutron scattering experiments as occurs in the case of a helimagnet,
although it can couple to the spin-spin correlations through a dynamic effect.
However, the spatial inhomogeneity complicates the calculation.
This remains an interesting problem which we will not pursue further 
in this work.

If $h_{so}\ge h_{so,L}$, then both instabilities are suppressed,
and thus in this regime the ground state is uniform.
In this case, the Goldstone spectrum has an
overall shift based on Eq.\eqref{eq:DM2D} as
\bea
\omega^{\prime 2}_l=\omega^2_l+ 2|f^a_1| \bar n h_{so}, \ \ \
\omega^{\prime 2}_t=\omega^2_t+ 2|f^a_1| \bar n h_{so}.
\eea
as depicted in Fig. \ref{fig:GSdispersion2}.
Then the spectra for the two transverse Goldstone modes exhibit
a roton like structure with a gap of
\bea
\Delta=\sqrt{ 2\bar n |f^a_1| (h_{so}-h_{so,T})}, 
\eea
at $q^\prime_c=\gamma_2 \bar n/(4\sqrt{2} \gamma_1)$.

\subsection{Goldstone modes in the 3D $\beta$-phase }
\label{sec:Goldstone}

For the 3D $\beta$-phase, we only consider the case of $l=1$ with the
ground state configuration of the $d$-vector as $\vec d(\vec k)\pp \vec k$.
The Goldstone modes behave similarly to the 2D case.
The Legendre conjugation operators of the
order parameter $n^{\mu,b}$ can be decomposed into the 
operators $O_{J,J_z} (J=0,1,2; J_z=-J,...J)$  as
eigen-operators of the  total angular momentum $\vec J=\vec L+\vec S$
in the $\beta$-phase.
The Higgs mode carries $J=0$ defined as
\bea
O_{hig}(\vec r)= \frac{1}{\sqrt 3}\delta_{\mu a} n^{\mu a}(\vec r).
\eea
Due to the broken relative spin-orbit symmetry, the relative spin-orbit 
rotations generate the three Goldstone modes $O_{x}$, $O_y$, and $O_z$
\bea
\hat O_{i}(\vec r)&=& \frac {1}{\sqrt 2} \epsilon_{i\mu a} n^{\mu,a}(\vec r),
\eea
which carry total angular momentum $J=1$.
We choose the propagation wavevector $\vec q$ along the $z$-axis. 
Due to presence of $\vec q$, only $J_z$ is conserved.
The relation between $O_{x}$, $O_y$ and $O_z$ and those in the helical basis
$O_{1,\pm1}$ and $O_{10}$ is
\bea
O_{1,\pm1}= \frac{1}{\sqrt 2} (O_{x}\pm i O_{y}), \ \ \ O_{1,0}=O_{z}.
\eea

In the low frequency and small wavevector regime as
$\omega, v_F q\ll \bar n \ll v_F k_F$, we can neglect the mixing 
between Goldstone modes and other massive modes.
Similarly to the case in 2D, we calculate the 
fluctuation kernels of the Goldstone modes as
\bea
L_{10,10}(q,\omega)&=& \kappa  q^2 -\frac{\omega^2}
{4\bar n^2 |f^a_1|}, \\
\label{eq:L10}
\hspace{-10mm} L_{1\pm1,1\pm1}(q,\omega)&=& \kappa  q^2
\pm\frac{N(0)}{18} \frac{q}{k_F} x -\frac{\omega^2} {4\bar n^2 |f^a_1|}.
\label{eq:L11}
\eea
For simplicity, we have neglected the anisotropy in the $q^2$ term 
in Eq. \eqref{eq:L11}.
Their spectra read
\bea
\omega_{j_z}^2=4\bar n^2 |F_1^a| \big ( \frac{\kappa q^2}{N(0)}
+j_z \frac{|q| x}{18 k_F} \big )~ (j_z=0,\pm 1). \ \ \
\eea
Because of the broken parity, the channel of $j_z=-1$ is unstable,
which leads to the Lifshiz-like instability as discussed
in the 2D $\beta$-phase.
 
Again this Lifshitz instability is due to the nontrivial $\gamma_2$
term in the G-L free energy of Eq.\eqref{eq:3Dinstab}.
To determine the coefficients of the gradient terms,
we linearize the $\gamma_2$ term around the saddle point,
define the deviation from the uniform mean field ansatz as
\bea
\vec n_1 &=& \sqrt{\bar n^2-\delta n_1^2} \hat e_x +\delta \vec n_1, \ \ \
\vec n_2 = \sqrt{\bar n^2- \delta n_2^2} \hat e_y +\delta \vec n_2,
\nonumber \\
\vec n_3 &=& \sqrt{\bar n^2- \delta n_3^2} \hat e_z +\delta \vec n_3,
\eea 
where $\delta \vec n_1 \perp \hat e_x$, $\delta \vec  n_2 \perp \hat e_y$,
and $\delta \vec  n_3 \perp \hat e_z$.
The contribution from the Goldstone modes becomes
\bea
F_{grad}(O)&=& \gamma^\prime_1 \{ (\partial_i O_x)^2 + (\partial_i O_y)^2
+(\partial_i O_z)^2 \}\nonumber\\
&-&  \frac{\gamma^\prime_2}{2} \bar n \epsilon_{ijk}  O_{i}\partial_j O_{k}
+ \gamma_2^\prime \sqrt 2 \bar n^2 \partial_i O_{i}.
\eea
Similarly to the 2D case, the values of $\gamma^\prime_{1}$ and $\gamma^\prime_2$ 
can be determined by matching the coefficients of the dispersion relation
in Eq.\eqref{eq:dispersion} as
\bea
\gamma^\prime_1\approx \kappa, \ \ \
\gamma^\prime_2=\frac{N(0)}{18 v_F k^2_F},
\label{eq:gamma3d}
\eea
where the difference among the three Frank constants is neglected.

Following the same procedure in the 2D case in Section \ref{subsec:lifshitz},
we study the instabilities of the longitudinal and transverse twists
in 3D $\beta$-phase
with an external spin-orbit field to pin the order parameter
\bea
V_{so}=-h^{3D}_{so} (n^{x,1}+n^{y,2}+n^{z,3}).
\eea 
After straightforward calculation, we find that longitudinal
twist occurs at the pitch wavevector $q_{L,3D}=\bar 
n \gamma_2/(2 \gamma_1)$ with a critical value of $h^{3D}_{so}$ field
to suppress the twist at $h^{3D}_{so,L}=\gamma_2^2 
\bar n^3/(4\gamma_1)$, and those of the transverse twist
are $q_{T,3D}=q_{L,3D}/2$ and $h^{3D}_{so,T}=h_{so,L}/2$.
Thus the conclusion is the same as in the 2D case that
the instability of the longitudinal twist is stronger than 
that of the transverse twist.
Again, if $h^{3D}_{so}\ge h^{3D}_{so,L}$, then the ground state is uniform,
and the spectrum of the two transverse Goldstone modes exhibits
a roton-like structure with a gap of
\bea
\Delta=\sqrt{ 2\bar n |f_1^a| (h_{so}-h_{so,T})}, 
\eea
located at $q_{T,3D}$.

\section{Magnetic field effects at zero temperature}
\label{sec:magnetic}

\begin{figure}
\psfrag{1}{$l=1$}
\psfrag{2}{$l=2$}
\psfrag{B}{$\vec B$}
\begin{center}
\includegraphics[width=0.4\textwidth]{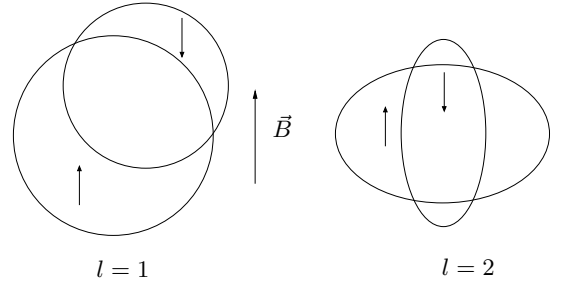}
\end{center}
\caption{The $\alpha$-phases in a $\vec B$ field ($l=1,2$). 
The larger Fermi surface has spin parallel to the $\vec B$ field.
The Fermi surface distortion is not pinned by the $\vec B$ field.
}\label{fig:Bfield1}
\end{figure}

\begin{figure}
\psfrag{B}{$\vec B$}
\psfrag{1}{$\vec n_1$}
\psfrag{2}{$\vec n_2$}
\begin{center}
\includegraphics[width=0.2\textwidth]{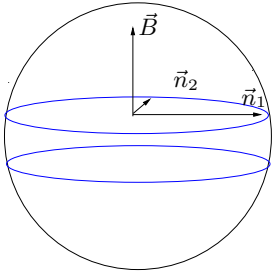}
\end{center}
\caption{The spin configuration on the Fermi surfaces in the
$\beta$-phase ($F^a_l$ channel). With the $\vec B$ field, the
large circle splits to two small circles
perpendicular to the $\vec B$ field with the development of a finite
magnetization.
The large (small) Fermi surface polarizes parallel (anti-parallel) to
the $\vec B$ field.
}\label{fig:Bfield2}
\end{figure}  

\begin{figure}
\psfrag{B}{$\vec B$}
\psfrag{x}{$\vec n_1$}
\psfrag{y}{$\vec n_2$}
\psfrag{z}{$\vec n_3$}
\begin{center}
\subfigure[]{\includegraphics[width=0.2\textwidth]{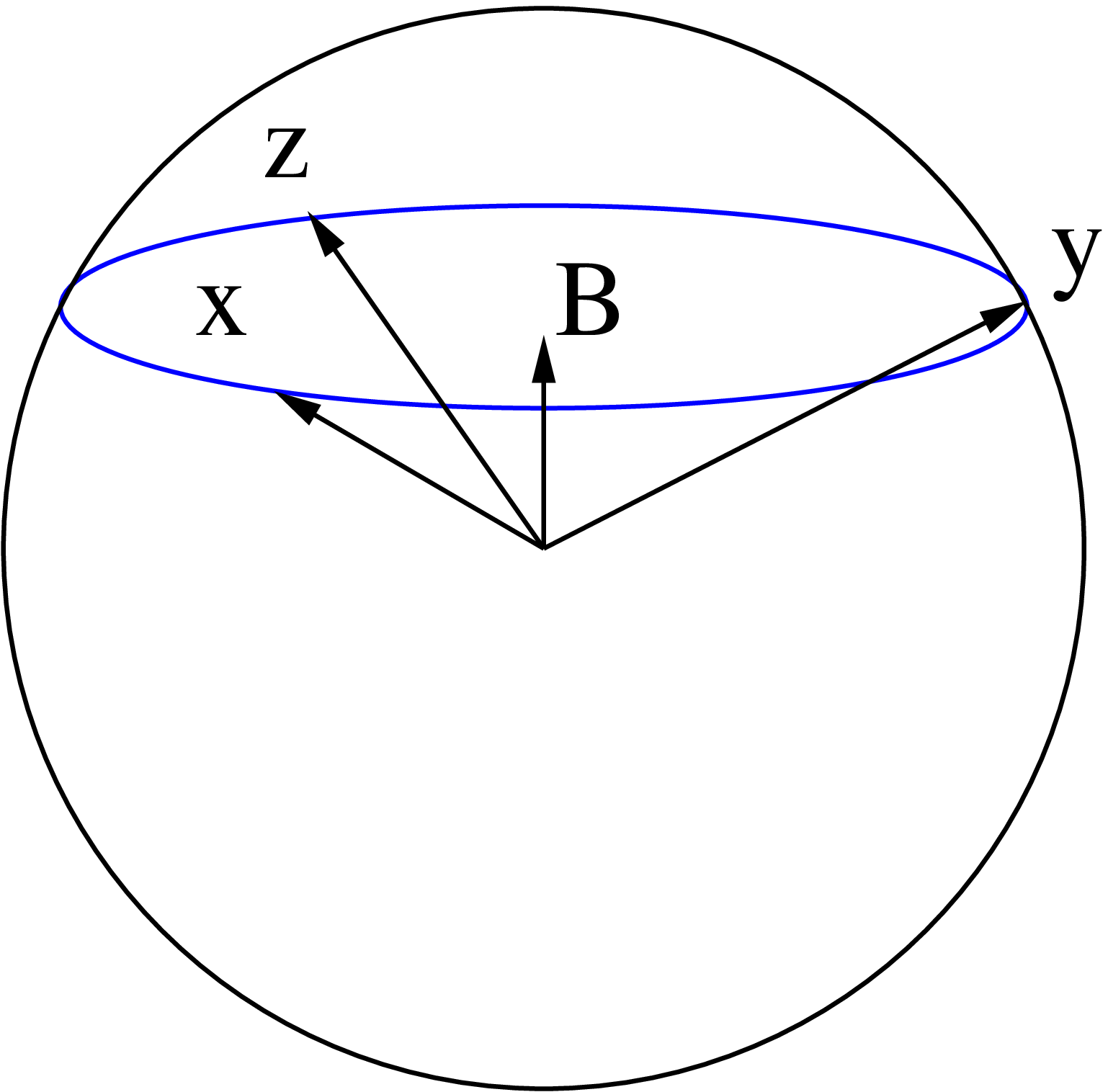}}
\subfigure[]{\includegraphics[width=0.2\textwidth]{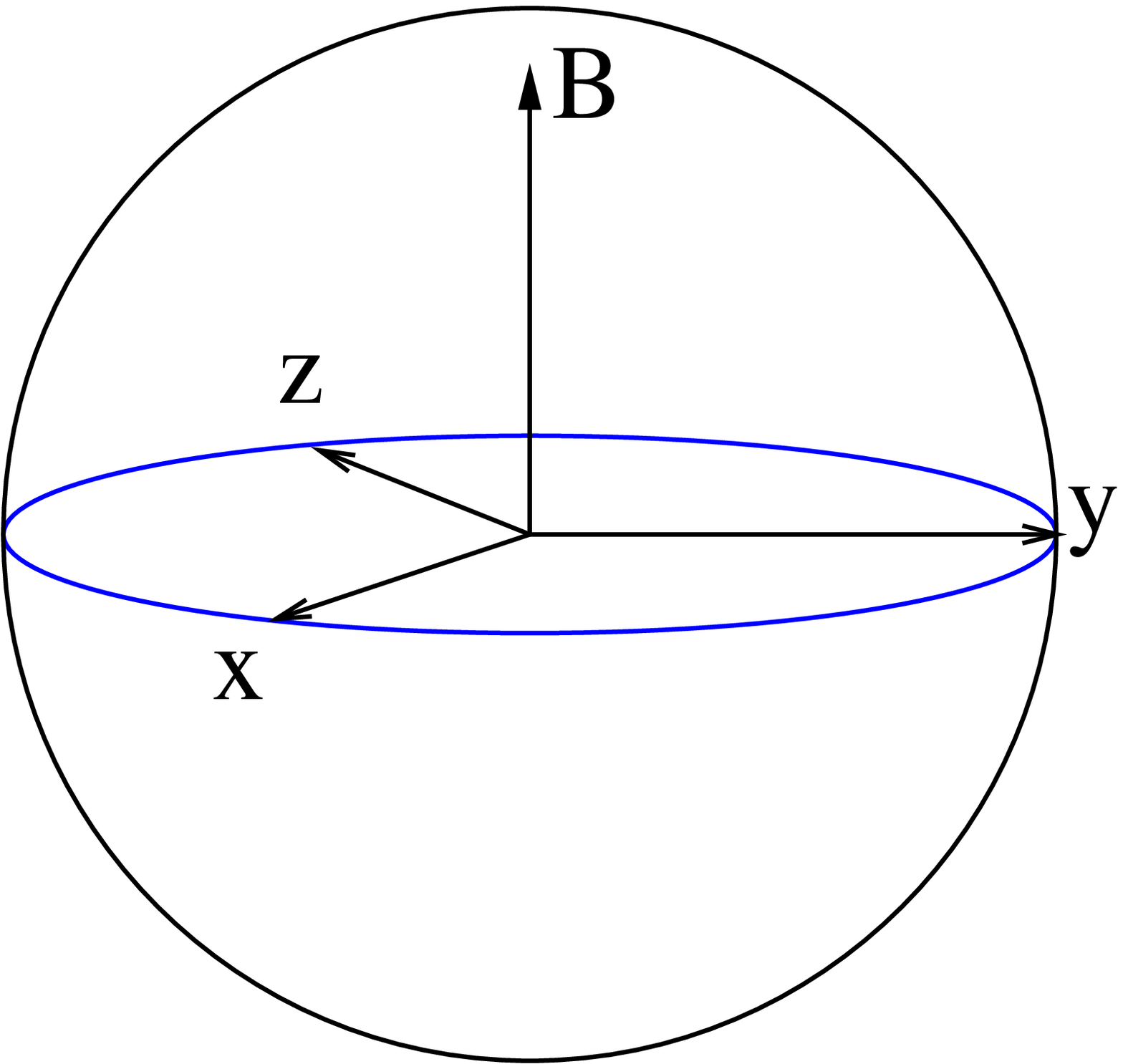}}
\end{center}
\caption{The order parameters in the 3D $\beta$-phase at $l=1$.
a) Configuration for $B^\prime>B$ as defined in Eq. \eqref{eq:BfieldBeta};
b) Configuration for $B_c>B>B^\prime$.
}\label{fig:Bfield3}
\end{figure} 

In this section, we discuss the magnetic field effect to the
order parameter configurations in the $\alpha$ and $\beta$-phases.
Due to the symmetry constraint, the $\vec B$ field does not couple to
the order parameter $n^{\mu a}$ linearly, and
the leading order coupling begins at the quadratic level as
\bea
\Delta F(\vec B, \vec n_b) &=&  (u+w)  g^2 B^2 \tr [n^T n] 
-w g^2 B^\mu B^\nu  n^{\mu b} n^{\nu b} \nonumber \\
&=&
u g^2  B^2 \sum_b  n_b^2
+ w g^2 \sum_b  (\vec B \times \vec n_b)^2,
\label{eq:Bfield}
\eea
with $g$ the gyromagnetic ratio.
The coefficients $u$ and $w$ can be determined from the microscopic
calculation by using the mean field approach as
\bea
u&=&4 v_1-2 v_2, \ \ \ w=4 v_2 ~(\mbox{at 2D}), \nn \\
u&=&6 v_1-3 v_2, \ \ \ w=5 v_2 ~(\mbox{at 3D and $l=1$}),
\eea
with $v_{1,2}$ defined in Eq.\eqref{eq:2Dcoeff} in 2D
and Eq.\eqref{eq:3Dcoeff} in 3D respectively.
$v_1$ is usually larger than $v_2$, thus $u$ is typically positive. 
This means that the external $\vec B$ field suppresses the
magnitude of order parameters as
\bea
\bar n(B)=\bar n(B=0)\sqrt{1-\frac{B^2}{B^2_c}}
\eea
in both $\alpha$ and $\beta$-phases, where
the critical value of $B_c$ is defined as
\bea
\frac{\mu_B B_c}{v_F k_F}=\frac{\sqrt{\frac{|r|}{u}}}{v_F k_F} 
=\frac{\bar n (0)}{k_F v_F}\ll 1.
\eea
This effect has been studied in Ref. \cite{varma2006}.

In the $\alpha$-phase, $w$ is positive.
Thus, the spin components of the order parameter of this phase, $\vec n_b$,
prefers to be parallel or anti-parallel to the magnetic field $\vec B$.
Consequently, 
the Fermi surface with  its spins polarized parallel to the magnetic field 
becomes larger in size than the Fermi surface for the spins pointing in the opposite direction. In the case of the nematic-spin-nematic phase, the $\alpha$-phase with angular momentum $l=2$, the main effect of the magnetic field is to break the symmetry by $\pi/2$ rotations, {\it i.e.\/} a spacial rotation followed by a spin reversal, while retaining the symmetry under rotations by $\pi$. Thus, the magnetic field induces a non-zero component of the charge nematic order parameter, the nematic in the spin-singlet channel. (Analogs of this effect hold in all angular momentum channels).
On the other hand, for a translation and rotationally invariant system, the Fermi surface distortion, {\it i.e.\/} orientation of the order parameter in real space, cannot be locked to the direction of 
the $\vec B$ field.
As a result, the Goldstone manifold becomes $[SO_S(2) \otimes
SO_L(2)] /SO_S(2)=SO_L(2)$.

The $B$ field in the $\beta$-phase also constrains the direction 
of the vectors $\vec n_b$. Since
$w$ is negative in the $\beta$-phase, $\vec B$ prefers to
be perpendicular to $\vec n_b$ vectors.
Thus in the 2D system, $\vec n_1$, $\vec n_2$, and $\vec B$ form 
a triad which can be either left-handed or right-handed
as depicted in Fig. \ref{fig:Bfield2}. 
The spin configuration on Fermi surfaces changes from a large circle
into two smaller circles with a net spin polarization along
the $\vec B$ field.
The triad still has the degrees of freedom to rotate around the
axis of $\vec B$, thus the Goldstone manifold at 2D is reduced
from $SO_{L+S}(3)$ to $SO_{L+S}(2)\times Z_2$. 
The $B$ field effect in 3D systems is more subtle.
At small values of $B$, $\vec n_1$, $\vec n_2$ and $\vec n_3$ can no longer for a triad with $\vec B$. Instead, $\vec n_1$, $\vec n_2$ and $\vec n_3$ for a distorted triad with the direction of the distortion
lying in the diagonal direction and parallel to $\vec B$, as  depicted in Fig. \ref{fig:Bfield3}. A.
As $B$ grows larger, the three vectors $\vec n_1$, $\vec n_2$ and $\vec n_3$ are pushed towards the
plane perpendicular to $\vec B$.
For $B$ larger than a value $B^\prime$, defined as
\bea
B^\prime=B_c\sqrt{\frac{v_2}{v_2+4 |w| v_1/u}},
\label{eq:BfieldBeta}
\eea  
$\vec n_1$, $\vec n_2$ and $\vec n_3$ become co-planar, and with a relative angle of
$2\pi/3$, {\it i.e.\/} an equilateral triangle,  as depicted in Fig. \ref{fig:Bfield3} B. 
If $B$ is further increased,  $\vec n_1$, $\vec n_2$ and $\vec n_3$ keep these directions
but their magnitudes continuously shrink to zero at $B=B_c$.

We next discuss the effect of magnetic field $B$ on the
Lifshitz-like instability in the 2D $\beta$-phase at $l=1$.
We assume that $\vec B$ is parallel to the $z$-direction.
As showed in Section \ref{subsec:lifshitz}, the instability of 
the longitudinal twist described in Eq. \eqref{eq:ltwist} is stronger 
than that of the transverse twist. Thus, we will focus here on the 
case of the longitudinal twist.
In this case, the order parameter $\vec n_{2}$ precesses around the
$\hat e_1$ axis. 
Thus, $\vec B$, $\vec n_1$ and $\vec n_2$ cannot form a fixed triad uniformly in space.
The free energy of Eq. \eqref{eq:GL2} now becomes:
\bea
V(n)&=&\gamma_1 \bar n^2 q^2 -\gamma_2 \bar n^3 q-w B^2 \bar n^2
(1+\cos^2 qx). ~~~~~
\eea
Hence, for $B$ larger than a critical value of $B_{cl}$,
\bea
B_l=\frac{B_c}{\sqrt{1+8 v_1 |w| \gamma_1/(u \gamma_2^2)}},
\eea
the Lifshitz instability of the $\beta$-phase to a phase with 
a longitudinal twist is suppressed.
The effects of an external $B$ field  in the 3D
$\beta$-phase are more complicated and will not be discussed here.

Finally, we note that the order parameter $n^{\mu a}$ can couple to 
the magnetic field $B$ linearly when other external fields are also present.
For $l=1$, a possible additional term 
in the free energy of the form
\bea
\Delta F(B,j)= \gamma_3  B^\mu j_a (\vec r)n^{\mu,a},
\eea
where $j_a$ is the $a$-th component of the  electric current.
To leading order, the mean-field value of the coupling
constant $\gamma_3$ is
\bea
\gamma_3=-\frac{\hbar^2 e \mu_B}{2 m^* v_F}
\frac{N(0)}{v_F k_F},
\eea
where $m^*$ is the effective mass and $e$ is the charge of
electrons. 
For $l=2$, a similar term can be constructed as
\bea
\Delta F(B, u)= \gamma_4  B^\mu u_a (\vec r) n^{\mu,a},
\eea
where $u_a$ is the strain field.
Due to these terms, in the $\alpha$-phase, 
the electric current and lattice strain can be used to lock the 
direction of the order parameter in real space in
the presence of an external magnetic field
at $l=1$ and $l=2$ respectively. 
In the $\beta$-phase, this term will distort the
round Fermi surfaces into two orthogonal ellipses but with
different volume.

\section{Spin current induced by a charge current in the $d$-wave
channel}\label{sec:spincurrent}
In the $d$-wave ($l=2$) case, the order parameters have
the structure of the spin-quadruple moments.
From the symmetry analysis, a spin current $J^{\mu, a}_s$ may be induced
by a charge current $J^a_c$ flowing through the system
where $\mu$ is the spin index, and $a, b$ are the spatial indices.
For simplicity, we only study the 2D $\alpha$ and $\beta$-phases.
In the standard quadruple notation, the order parameters 
$n^{\mu,1}$ and $n^{\mu,2}$ can be represented as
$n^{\mu,1}= n^{\mu,xx}-n^{\mu,yy}$, and $n^{\mu,2}=2 n^{\mu,xy}
=2 n^{\mu,yx}$.
Then we write the formula as
\bea
J^{\mu, a}_s= g \bar n^{\mu, ab} J_{c}^b,
\eea
where the matrix $\bar n^{\mu, ab}$ is related to the order 
parameter $\bar n^{\mu,a}$
as
\bea
\bar n^{\mu,ab}= 2 \left(\begin{array}{c c}
\bar n^{\mu,xx} & \bar n^{\mu,xy} \\
\bar n^{\mu,yx} & \bar n^{\mu,yy}
\end{array}\right)
=
\left(\begin{array}{c c}
\bar n^{\mu,1} & \bar n^{\mu,2} \\
\bar n^{\mu,2} & -\bar n^{\mu,1}
\end{array}\right).
\eea
By the standard the linear response theory, the coefficient $g$ can
be calculated as
\bea
g=\frac{(3-2a)\pi}{e k_F v_F |F^a_2|}.
\eea

In the $\alpha$-phase, it is convenient to choose the direction of the 
axes of the reference frame $x$ and $y$ along the major and minor axes of
the distorted Fermi surfaces, and assume spin quantization along 
$z$-axis, so that $\bar n^{\mu, ab}
=\bar n \hat e_z \mbox{diag}\{1, -1\}$.
A charge current $ J^a_c$ running along the major and minor axes
induces a spin current $J^{\mu,a}_s$ flowing in the same 
(or opposite) direction.
But for the general direction of $J^a_c$, the induced spin 
current $J^{\mu, a}_s$ flows with an angle with $J^a_c$.
We denote the azimuthal angle between the charge current $J^a_c$
and the $x$-axis as $\phi$.
Then the angle between $J^{a}_c$ and $J^{\mu, a}_c$
reads $2\phi$ or $\pi-2\phi$ depending on the sign of $g$.
The nature of the induced spin current here is different
from that of the spin-Hall effect in semi-conductors with
SO coupling.
In that case, the spin-Hall current always flows 
perpendicular to the electric field, and the spin
Hall conductance is invariant under time-reversal transformation.
Here, because of the anisotropy of the Fermi surfaces,
the spin current is perpendicular to the charge current, 
only if the  charge current  flows  along the diagonal 
direction ($\phi=\pm \pi/4, \pm 3\pi/4$).
On the other hand, the $d$-wave phases break time reversal
symmetry, thus the induced spin current is not dissipationless.

In the $\beta$-phase, without loss of generality, we can take 
order parameter configuration as in Fig. \ref{fig:vortex2}. a,
i.e., 
\bea
n^{\mu, ab}= \bar n \left(\begin{array}{c c}
\hat e_x  & \hat e_y  \\
\hat e_y  & -\bar e_x
\end{array}\right),
\eea
where $\hat e_{x,y}$ denote the spin direction.
We assume that the charge current $J^{a}_c$ flows along the $x$ direction.
Then two spin currents polarizing along orthogonal directions,
are induced with the same magnitude.
The spin current flowing along the $x$-direction polarizes
along $\hat e_x$, while that flowing along the $y$-direction
polarizes along $\hat e_y$.
If we measure the spin current along the spatial direction
with the azimuthal angle $\phi$ respect to the $x$-axis,
the induced spin current along this direction
polarizes along the direction of
$\cos \phi ~\hat {e}_x +\sin \phi ~\hat {e}_y$.
Because in the $\beta$-phase, there is an induced SO coupling,
spin and orbital angular momenta will not be 
preserved separately. 
As a result, it is impossible in the $\beta$-phase to describe a spin
current by two separated indices (a spatial index and a spin one). The spatial 
degrees of freedom and the spin ones must be mixed together.

Finally, in a real material, there would always be some SO coupling. 
With even an infinitesimal SO coupling, 
a charge current flowing inside the system can remove the degeneracy
of the ground states in the ordered phase. In other words, in the presence of 
explicit SO interactions, a charge current can pin down the direction 
of the order parameter. Therefore, the relative
angle between the order parameter and the charge current is not arbitrary. 
As a result, to be able to adjust the angle between the charge current and the order
parameter as we mentioned above, some other mechanism is necessary
to pin down the order parameter such that the order parameter will not
rotate when we rotate the direction of the charge current. For example, an in-plane 
magnetic field or a lattice potential as background can do the job.

\section{Possible experimental evidence for these phases}
\label{sec:experiment}

At present time, we are not aware of any conclusive experimental evidence
for a spin triplet channel Pomeranchuk instability.
However, the $\alpha$ and $\beta$-phases presented here are just
a natural generalization of ferromagnetism to
higher partial wave channels.
Taking into account the existence of the $p$-wave Cooper pairing phase
in the $^3$He systems\cite{leggett1975} and the strong evidence for its existence 
in the ruthenate
compound Sr$_2$RuO$_4$ \cite{mackenzie2003},
we believe that there is a strong possibility to find these phases
in the near future. Basically, the driving force for behind the Pomeranchuk instabilities in the spin 
triplet channels is still the exchange interaction among electrons, 
which shares the same origin as in ferromagnetism. Although the weak coupling analysis we have used here may not apply to the materials of interest, as many of them are strongly correlated systems, many of the symmetry issues will be the same as the ones we have discussed here, with the exception of the role of lattice effects which we have not addressed in detail and which may play a significant role, {\it i.e.\/} by gapping-out many of the Goldstone modes associated with the continuous rotational symmetry of the models that we have discussed. Nevertheless that GL free energies will have much of the same form even if the actual coefficients may be different, since we typically need a strong enough exchange interaction in a non $s$-wave
channel.
In the following, we will summarize a number of known experimental systems
(and numerical) which suggest possible directions to search for 
the $\alpha$ and $\beta$-phases.

\paragraph{$^3$He}:
The spin exchange interaction in the Fermi liquid state of 
$^3$He is very strong, as exhibited in the low frequency
 paramagnon modes \cite{leggett1975}.
in this system, the spin fluctuations are known to mediate the $p$-wave Cooper pairing.
The Landau parameter $F^a_1$ in $^3$He was determined to be 
negative from various experiments
\cite{leggett1970,corruccini1971,osheroff1977,greywall1983}, including
the normal-state spin diffusion constant, spin-wave spectrum,
and the temperature dependence of the specific heat. It
varies from around $-0.5$ to $-1.2$ with increasing pressures to
the melting point.
Although $F^a_1$ is not negatively large enough to pass the
critical point, we expect that reasonably strong fluctuation effects
exist.

\paragraph{URu$_2$Si$_2$}:
The heavy fermion compound URu$_2$Si$_2$ undergoes a phase transition
at 17K. 
The tiny antiferromagnetic moment developed in the low 
temperature phase can not explain the large entropy loss. 
About $40\%$ density of state density is lost
at low temperatures.
Currently the low temperature phase is believed to be characterized by
an unknown `hidden' order parameter \cite{tripathi2005}.
An important experimental result of nuclear magnetic resonance (NMR) 
\cite{bernal2001} shows the broadening of the line shape below T$_c$.
This implies the appearance of a random magnetic field in the
hidden ordered phase.
Recently, Varma {\it et al.} proposed the $p$-wave $\alpha$-phase
as the hidden ordered phase \cite{varma2005,varma2006}.
They fit reasonably well the specific heat jump, and more 
importantly the jump of the non-linear spin susceptibility $\chi_3$
at the transition.
The origin of the random field in the NMR experiment is explained 
by the spin moment induced by disorder in the $p$-wave $\alpha$-phase. 
However, the $\alpha$-phase still has Fermi surfaces, thus it
is difficult to explain the large loss of density of states.
Further, the $\alpha$-phase is time reversal even, thus its
coupling to spin moment must involve $B$ field.
In the NMR experiment, an external $B$ field is indeed added.
It would be interesting to check whether the line-shape
broadening is correlated with the magnitude of $B$.

\paragraph{Sr$_3$Ru$_2$O$_7$}:
The bilayer ruthenate compound Sr$_3$Ru$_2$O$_7$ develops a metamagnetic 
transition in an applied magnetic field $B$ perpendicular to the  $c$-axis.
In very pure samples, for $B$ from 7.8 $T$ to 8.1 $T$, the resistivity
measurements show a strong enhancement below 1.1 K \cite{grigera2004}.
Transport measurements in tilted magnetic fields, with a finite component of the $B$ field in the $ab$ plane,  shows evidence for a strong in-plane temperature-dependent anisotropy of the resistivity tensor, which is suppressed at larger in-plane fields\cite{grigera2004,mackenzie-comm}.
This effect is interpreted as a nematic transition for the Fermi
surface of the majority spin component \cite{grigera2004}. This result suggests a state which
is a superposition of both a charge nematic and a nematic-spin-nematic state.
On the square lattice, the $d_{x^2-y^2}$ distortion pattern is
more favorable than that of $d_{xy}$.
Thus the transition should be Ising-like. 
In the presence of SO coupling, while preserving both parity and TR symmetries,
the magnetic field can couple to the $d_{x^2-y^2}$ channel order parameter through terms in the free energy of the form
\bea
(B_x^2-B_y^2) B_z n_{z,1},
\eea
which is cubic in $B$, and
\be
\vec B \cdot \vec n_a \; n_a
\ee
(where $n_a$ is the charge nematic order parameter), which is linear in $B$.
Thus, an in-plane $B$ field can lock the orientation of
the nematic-spin-nematic order parameter. 
This effect is more pronounce if the system is in a charge nematic phase.

\paragraph{2DEG in large magnetic fields}:
Currently, the strongest experimental evidence for a charge nematic (fully polarized) state is in the case of a 2DEG in a large perpendicular magnetic field\cite{lilly1999,du1999,cooper02}. In the second and higher Landau levels, a huge and strongly temperature-dependent resistance anisotropy is seen in ultra-high mobility samples, for filling factors near the middle of the partially filled Landau level. In this regime, the I-V curves are clearly linear at low bias. No evidence is seen of a threshold voltage or of broad band noise, both of which should be present if the 2DEG would be in a stripe state, which is favored by Hartree-Fock calculations. Both effects are seen in nearby reentrant integer-Hall states. Thus, the simplest interpretation of the experiments is that the ground state is a polarized charge nematic\cite{fradkin99,fradkin2000}. There is still  a poorly understood alternation effect: the strength of the anisotropic resistance appears to alternate between the fully polarized state and the state with partial polarization. Although this effect could be explained in terms of microscopic calculations of the order parameter, it is still possible that the latter may suggest some form of partially polarized nematic-spin-nematic order. there are no reliable calculations of Landau parameters in these compressible phases.

\paragraph{The 2DEG at zero magnetic field}:
The 2DEG at low densities is a strongly coupled system and much work has been done on this system in the context of its apparent metal-insulator transition. What interests us is the possibility that this system may have phases of the type discussed here. (The possibility of non-uniform ``micro-emulsion'' phases in the 2DEG was proposed recently by Jamei and collaborators\cite{jamei05}.)
A numerical evaluation to the Landau parameter $F_1^a$ in 2D, performed
by Kwon et al, \cite{kwon1994} by using variational quantum
Monte-Carlo, found that $F^a_1$ is negative and decreasing 
from $-0.19$ at $r_s=1$ to $-0.27$ at $r_s=5$.
On the other hand, Chen et al.  \cite{chen1999,chen2000}
investigated the many-body renormalization effect to the
Rashba SO coupling due to the exchange interaction in the $F^a_1$ channel.
They found the renormalized SO coupling is amplified significantly
at large $r_s$ by using a local field approximation.
More numerical work to check whether Pomeranchuk instabilities
can occur in this system would be desirable.

\paragraph{Ultra-cold atomic gases with a $p$-wave Feshbach resonance}:
Another type of strongly interacting system is cold atoms with 
Feshbach resonances.
Recently, a inter-species $p$-wave Feshbach resonance has been 
experimentally studied by using the two component $^6$Li atoms
\cite{zhang2004}.
In the regime of positive scattering length, close to the resonance
the Landau parameter of $F^a_1$ should be negative and large in magnitude. Thus, this system would appear to be a good candidate to observe these phases.
However, since the $p$-wave Feshbach resonance is subject to a
large loss-rate of particles, it is not clear whether it would be possible to use this approach to observe a stable
system with a Pomeranchuk instability near the resonance.

\paragraph{How to detect these phases}:
We also propose several experimental methods to detect the $\alpha$
and $\beta$-phases (see also the discussion in Ref.\cite{Podolsky2005}). For the case of the spatially anisotropic $\alpha$-phases, evidence for strongly temperature dependent anisotropy in the transport properties (as well as the tunability of this effect by either external in-plane magnetic fields and/or uniaxial stress)  as seen in Sr$_3$Ru$_2$O$_7$ and in the 2DEG can provide direct evidence for the spacial nematic nature of these phases. Spatially nematic phases exhibit anisotropic transport properties even in a single-domain sample\cite{oganesyan2001}. More difficult is to determine their spin structure.
Because no magnetic moments appear in both phases, elastic neutron
scattering does not exhibit the regular Bragg peaks.
Since the Goldstone modes are combined spin and orbital excitations, they can not be
directly measured through neutron scattering.
Nevertheless, in the ordered phase, as we have discussed spin-spin
correlation function couples to  the Goldstone modes, and develops a characteristic resonance
structure, which should be accessible to inelastic neutron scattering.
An experimental detection of this resonance and its appearance 
or disappearance in the ordered or disordered phases can justify
the existence of these phases.
On the other hand,  Fermi surface configurations and single particle
spectra in the $\alpha$ and $\beta$-phases are different from
 the normal state Fermi liquids.
If the angle resolved photon emission spectroscopy (ARPES) experiment
can be performed, it can readily tell these phases.
In the $\beta$-phase, the order parameter is similar to the Rashba SO coupling,
the method to detect the Rashba coupling can be applied here.
For example, from the beat pattern of the Shubnikov-De Hass oscillations
of the $\rho(B)$, we can determine the spin-splitting of two helicity bands.
The asymmetry of the confining potential certainly will also contribute
some part to the final spin-orbit coupling.
But, when the dynamically generated part dominates, it will not
sensitive to the asymmetry of the confining potential.

\section{Conclusions}
\label{sec:conclusion}

In summary, we have studied the Pomeranchuk instability involving spin
in the high orbital partial wave channels.
GL free energies are construct to understand the ordered phase
patterns after the instabilities take place.
The ordered phases can be classified into $\alpha$ and $\beta$-phases 
as an analogy to the superfluid $^3$He A and B phases.
Both phases are characterized by a certain type of effective SO coupling,
gives rise a mechanism to generate SO couplings in a non-relativistic
systems. In the $\alpha$-phase, the Fermi surfaces exhibit an anisotropic 
distortion, while those in the $\beta$-phase still keep the circular 
or spherical shapes undistorted.
We further analyze the collective modes in the ordered phases
at the RPA level.
Similarly to the Pomeranchuk instability in the spin-singlet density channel,
the density channel Goldstone modes in the $\alpha$-phase also shows
anisotropic overdamping, except along some specific symmetry-determined directions.
The spin channel Goldstone modes are found to exhibit nearly isotropic linear dispersion
relations at small propagating wave vectors.
The Goldstone modes in the $\beta$-phase are relative spin-orbit
rotation modes with linear dispersion relation at $l\ge 2$.
The spin-wave modes in both ordered $\alpha$ and $\beta$-phases
couples to the Goldstone modes, which thus develop characteristic resonance
peaks, that can be observed in inelastic neutron scattering experiments.
The $p$-wave channel is special in that the $\beta$-phase can
develop a spontaneous chiral Lifshitz instability in the
originally nonchiral systems.
The GL analysis was performed to obtain the twist pattern in the ground state.
We also review the current experiment status for searching these
instabilities in various systems, including $^3$He, the heavy fermion compound URu$_2$Si$_2$, the bilayer
ruthenate Sr$_3$Ru$_2$O$_7$, 2D electron gases, and $p$-wave
Feshbach resonances with cold fermionic atoms.
The Sr$_3$Ru$_2$O$_7$ seems the most promising systems to
exhibit such an instability in the 2D $d$-wave channel.
However, investigation on the SO coupling effect is needed
to understand the suppression of the resistivity anomaly
due to the in-plane field.

There are still many important properties of the spin triplet
Pomeranchuk instabilities yet to be explored.
In the paper, we did not discuss the behavior of the fermionic degrees of freedom, which are expected to be strongly anomalous.
Generally speaking, the overdamped density channel Goldstone modes
in the $\alpha$-phase strongly couple to the fermions, which is expected to lead to a
non-Fermi liquid behavior as in the case of the density channel
Pomeranchuk instabilities \cite{oganesyan2001,lawler2006}.
However, the Goldstone modes in the $\beta$-phase is not damped
as $l\ge 2$, thus similarly to the case of  itinerant ferromagnets,
the $\beta$-phase remains a Fermi liquid.
The $p$-wave channel in particularly interesting.
We have shown in Eq. \eqref{eq:gaugefm}, that $p$-wave 
paramagnon fluctuations couple to fermions as an $SU(2)$
gauge field.
The linear derivative terms in the G-L free energy also
become relevant in the finite temperature non-Gaussian regime.
The Hertz-Mills type critical theory for the $F^a_l$ contains
new features compared to the ferromagnetic ones.
We defer to a future publication to address the above interesting
questions.

\begin{acknowledgments}
We thank L. Balents, M. Beasley, S. L. Cooper, S. Das Sarma, M. P. A. Fisher, S. A. Grigera, S. A. Kivelson, H. Y. Kee, M. J. Lawler,
A. W. W. Ludwig, A. P. Mackenzie, C. M. Varma, and J. Zaanen for helpful discussions. 
CW thanks A. L. Fetter and Y. B. Kim for their pointing
out the Lifshitz instability in the $\beta$-phase. This work was supported in part by the National Science Foundation through the grants PHY99-07949 at the Kavli Institute for Theoretical Physics (C.W.), DMR-04-42537 at the University of Illinois (E.F.), DMR-0342832 at Stanford University (S.C.Z.), and by the Department of Energy, Office of Basic Energy Sciences,  under the contracts  DE-FG02-91ER45439 at the Frederick Seitz Materials Research Laboratory of the University of Illinois (E.F. and K.S.), and  DE-AC03-76SF00515 at Stanford University (S.C.Z.). 

\end{acknowledgments}

\begin{appendix}
\section{Landau interaction parameters}
\label{app:landau-parameters}

Landau-Fermi liquid theory is characterized by the interaction functions
which describe the forward scattering process between quasi-particles as
\bea
f_{\alpha\beta,\gamma\delta}(\vec p, \vec p^\prime)
= f^s(\vec p, \vec p^\prime)+ f^a(\vec p, \vec p^\prime)
\vec{\sigma}_{\alpha\beta}\cdot \vec{\sigma}_{\gamma\delta},
\eea
where $\vec p$ and $\vec p^\prime$ lie close to the Fermi surface.
The expressions of $f^s$ and $f^a$ can be obtained through
a general microscopic two-body $SU(2)$ invariant interaction 
\bea
V(\vec r_1, \vec r_2)= V_{c} (\vec r_1-\vec r_2)
+V_{s} (\vec r_1-\vec r_2) \vec S_1 \cdot \vec S_2,
\eea
where the $V_c$ and $V_s$ are  the spin-independent and dependent
parts, respectively.
At the Hartree-Fock level, $f^{s,a}(p,p^\prime)$ are
\bea
f^s(\vec p, \vec p^\prime)&=&V_c(0)-\frac{1}{2}V_c(\vec p- \vec p^\prime)
-\frac{3}{8} V_s(\vec p- \vec p^\prime) ,\nonumber\\
f^a(\vec p, \vec p^\prime)&=&-\frac{1}{2}V_c(\vec p-\vec p^\prime)
+\frac{1}{4} V_s(0) +\frac{1}{8} V_s(\vec p-\vec p^\prime).
\nonumber \\
\eea

$f^{s,a}(\vec p,\vec p^\prime)$ can be further decomposed into
different orbital angular momentum channels as
\bea
f^{s,a}_l=\left\{
\begin{array}{c}
\int^1_{-1} d \cos \theta~ f^{s,a}(\hat p\cdot \hat p^\prime)
P_l(\hat p \cdot \hat p^\prime) \ \ \ \mbox{in 3D}, \\
\int \frac{d \phi}{2\pi}  f^{s,a}(\hat p\cdot \hat p^\prime)
\cos l \phi \ \ \  \mbox{in 2D}
\end{array}
\right.
\eea
where $P_l$ is the $l$th-order Legendre polynomial.
For each channel of $F^{s,a}_l$,
Landau-Pomeranchuk (LP) instability \cite{pomeranchuk1959} occurs  at 
\bea
F^{s,a}_l= N(0) f^{s,a}_l \left \{ \begin{array}{c}
<-(2 l+1) \ \ \ \mbox{in 3D}, \\ 
< -(2- \delta_{l,0}) \ \ \  \mbox{in 2D},
\end{array}
\right.
\eea
where $N(0)$ is the density of states at the Fermi energy.

\section{the Goldstone modes of the spin oscillation in the alpha phase}
\label{app:gsspinealpha}
In this section, we calculate the spin channel Goldstone modes in the 2D 
$\alpha$-phases at small wavevector $v_F q/\bar n \ll 1$ and low frequency
$\omega/\bar n \ll 1$.
The expression for the dispersion of $L_{ x\pm iy,1}(\vec q,\omega)$ is
\bea
&&L_{x+iy,1}(\vec q, \omega)= \kappa  q^2 +\frac{1}{|f^a_1|}
+2 \int \frac{d^2 k}{(2\pi)^2} \cos^2 l \theta_k\nn \\
&\times&\frac{ n_f [\xi_\downarrow(\vec k-\frac{\vec q}{2})]
-n_f [\xi_\uparrow(\vec k+\frac{\vec q}{2})]}{\omega+i\eta
+\xi_\downarrow(\vec k-\frac{\vec q}{2})-\xi_\uparrow(\vec k+\frac{\vec q}{2})},\nn \\
\label{eq:gsspinealpha}
\eea
where 
$\xi_{\uparrow,\downarrow}(k)=\epsilon(k)-\mu\mp \bar n \cos l \theta_k$.
Following the procedure in Ref. \cite{oganesyan2001},
we separate Eq. \eqref{eq:gsspinealpha} into a static part $L_{x+iy,1}
(\vec q,\omega)$ and a dynamic part $M_{x+iy,1}(\vec q,\omega)$
as 
\bea
L_{x+iy,1}(\vec q, \omega)=L_{x+iy,1}(\vec q, 0)+ 
M_{x+iy,1}(\vec q,\omega).
\eea
At $\vec q=0, \omega=0$, from the self-consistent equation 
Eq. \eqref{eq:sfconalpha}, the integral cancels the constant term
$\frac{1}{|f_1^a|}$  as required by Goldstone theorem.
The detailed form of the static part at small but nonzero
$\vec q$ is difficult to evaluate due to anisotropic Fermi surfaces.
Because of the breaking of parity in the $\alpha$-phase, it seems
that the leading order contribution should be linear to $q$.
However, from the Ginzburg-Landau analysis in Sec. \ref{subsec:2dalpha},
the linear derivative term in Eq. \eqref{eq:GL2} does not contribute to
the coupling among Goldstone modes, i.e., the uniform ground state is stable
in contrast to the case in the $\beta$-phase.
As a result, the dependence on $\vec q$ should start from the quadratic order,
bringing a correction to the coefficient $\kappa$.
For simplicity, we neglect this correction for it does not cause qualitatively
different result.
Thus, we arrive at
\bea
L_{x+iy,1}(\vec q, \omega=0)=\kappa q^2.
\eea

The dynamic part, $M_{x+iy,1}(\vec q, \omega)$, can be expressed as
\bea
M_{x+iy,1}&=&
-2\int \frac{d^2 k}{(2\pi)^2} \frac{ \cos^2 l \theta_k~ \omega}
{\omega+i\eta+\xi_\downarrow(\vec k-\frac{\vec q}{2})-\xi_\uparrow(\vec k+\frac{\vec q}{2})} \nn \\
&\times&\frac{ n_f [\xi_\downarrow(\vec k-\frac{\vec q}{2})]-n_f [\xi_\uparrow(\vec k+\frac{\vec q}{2})]}{
\xi_\downarrow(\vec k-\frac{\vec q}{2})-\xi_\uparrow(\vec k+\frac{\vec q}{2})}.
\label{eq:dynamic}
\eea
To evaluate this integral, we  make several simplifications:
the non-linear part in $\epsilon(\vec k)$ is neglected and
the linear order in $\vec q$ is kept in the denominator.
We arrive at
\begin{widetext}
\bea
&&\int \frac{d \theta_k}{2\pi}  
\frac{\cos^2 l \theta_k}{\omega+i\eta
+2\bar n \cos l \theta_k - q v_F \cos (\theta_k-\phi)}
\frac{-\omega}{2\bar n \cos l \theta_k - q v_F \cos (\theta_k-\phi)}
\int\frac{k dk}{2\pi} (n_f [\xi_\downarrow(\vec k-\frac{\vec q}{2})]-n_f
[\xi_\uparrow(\vec k+\frac{\vec q}{2})]),\nn \\
&=&\int \frac{d \theta_k}{2\pi}  
\frac{-\omega~\cos^2 l \theta_k}{\omega+i\eta
+2\bar n \cos l \theta_k - q v_F \cos (\theta_k-\phi)}
\frac{k_2^2-k_1^2}
{2\pi [2\bar n \cos l \theta_k - q v_F \cos (\theta_k-\phi)]}, \nn \\
\eea
\end{widetext}
where $k_1$ and $k_2$ satisfy $n_f [\xi_\uparrow(\vec k_1+\vec q/2)]=0$ and 
$n_f [\xi_\downarrow(\vec k_2-\vec q/2)]=0$ respectively,
and $\phi$ is the azimuthal angle of $\vec q$.
$k_1$ and $k_2$ can be approximated by
\bea
k_{1,2}&=&k_F (1-\frac{x^2}{4}\pm(\frac{\bar n \cos l \theta_k}
{v_F}
-\frac{q}{2}\cos (\theta_k-\phi))\nn\\ 
&+&O(q^2). 
\eea

We now transfer the integral over $\theta_k$ to an integral
over $z=\exp(i \theta_k)$ and define  the density of 
the states at chemical potential in the ordered state as
\be
N_<(0)=\frac{k_F(1-x^2/4)}{v_F\pi},
\ee
The integral above now reads as:
\begin{widetext}
\bea
&&\frac{N_<(0)}{2\pi i} \oint\frac{d z}{z}(\frac{z^l+z^{-l}}{2})^2 
\frac{\omega}
{\omega+i\eta
+2\bar n \frac{z^l+z^{-l}}{2} 
-\frac{q v_F}{2} \left(z e^{-i \phi}+z^{-1}e^{i \phi}\right)
+O(q^2)
}
\nonumber\\
&\approx& \frac{N_<(0)\omega}{4}
\sum_{|z|<1} \textrm{Res} \left\{
\frac{(z^l+z^{-l})^2}{z\left[2 \omega+2 i\eta
+2\bar n (z^l+z^{-l}) 
-q v_F (z e^{-i \phi}+z^{-1}e^{i \phi})\right]}\right\}.
\eea
\end{widetext}
This integral can be calculated by evaluating the residues at poles 
inside the unit circle. 
There is one pole at $0$, one pole at $\infty$, and $2l$ poles,  
from the solutions of the equation: 
\bea
2\omega+2i\eta+2\bar n (z^l+z^{-l})
-q v_F(z e^{-i \phi}+z^{-1}e^{i \phi})=0.\nonumber \\
&&
\label{eq:pole}
\eea
The pole at $z=\infty$ is not 
inside the unit circle and does not contribute to the integral.

The pole at $z=0$ has different behavior for different values of $l$. 
The residue is  $-\frac{2\omega}{(2 \bar n -e^{i \phi} q v_F)^2}$
for $l=1$,  $-\frac{4 \bar n \omega-e^{2 i\phi}q^2 v_F^2}{8\bar n^3}$
for $l=2$, and $-\frac{\omega}{2\bar n^2}$ for all $l>2$. 
To leading order, all the residues  become
$-\frac{\omega}{2\bar n^2}$ at $l\ge 1$.
Next we discuss the poles at the solutions of Eq. \eqref{eq:pole}.
For $l\ge 1$, not all of these poles are located
inside the unit circle.
However, we will not bother to tell which poles are inside the unit
circle because we can show that these poles at most only gives 
negligible higher order terms.
In the limit of small $q$ and $\omega$,
Eq. \eqref{eq:pole} can be solved perturbatively as a power series
of $q$ and $\omega$ as
$z_m=\exp(\frac{i (2 m-1) \pi}{2 l})+O(q)+O(\omega)$, where 
$m=1,2,\dots,2 l$.
To the leading order, this type of poles are all simple poles, and
the residue of  $1/\left(2 \omega+2 i\eta+2\bar n (z^l+z^{-l}) 
-q v_F (z e^{-i \phi}+z^{-1}e^{i \phi})\right)$ is at the order of
$O(\frac{1}{2 \bar n})$. Therefore, the contribution from the pole
$z_m$ is:
\begin{widetext}
\bea
&&\frac{N_<(0) \omega}{4}\mbox{Res}(\frac{1}{z}
\frac{(z^l+z^{-l})^2}{2 \omega+2 i\eta
+2\bar n (z^l+z^{-l}) 
-q v_F (z e^{-i \phi}+z^{-1}e^{i \phi})})_{z=z_m}
\nonumber\\
&\approx&\frac{N_<(0) \omega}{2}\frac{(2 \omega+2 i\eta
-q v_F (z_m e^{-i \phi}+z_m^{-1}e^{i \phi}))^2}
{z_m(2 \bar n)^2} \frac{1}{4l \bar n z_m^{l-1} }
\approx O(\frac{\omega (\omega+q)^2}{\bar n^3}),
\eea
\end{widetext}
which is negligible to order  of
$O(\omega^2/\bar n^2)$ and $O(q^2/\bar n^2)$.

In short, for $l\ge 1$, only the pole at $z=0$ contributes to the
integral.
The result of the fluctuation kernel is
\bea
L_{x+iy,1}(\vec q, \omega)=\kappa q^2
-\frac{N_<(0)}{4\bar n^2}\omega^2.
\eea
From the self-consistent equation we find that $N_<(0)=\frac{2}{|f^a_l|}$.
Therefore,  the spin channel Goldstone mode reads
\bea
L_{x+iy,1}(\vec q,\omega)&=&\kappa q^2-\frac{\omega^2 }{2\bar n^2 |f^a_l(0)|}
\nn \\
&=&\kappa q^2-\frac{N(0)}{2 |F^a_l|} \frac{\omega^2 }{\bar n^2},
\eea
at $l\ge 1$.

\end{appendix}


\end{document}